\documentclass[fleqn]{llncs} 
\usepackage[hidelinks]{hyperref}
\usepackage{amsmath} 
\usepackage{version} 
\usepackage{ipapram} 

\hypersetup{colorlinks=true,linkcolor=blue,citecolor=blue,urlcolor=blue}

\title{Imperative Process Algebra \\ and Models of Computation}
\author{C.A. Middelburg \\
        {\small ORCID: \url{https://orcid.org/0000-0002-8725-0197}}}
\institute{Informatics Institute, Faculty of Science, University of
           Amsterdam \\
           Science Park~904, 1098~XH Amsterdam, the Netherlands \\
           \email{C.A.Middelburg@uva.nl}}
  
\begin{document}
\maketitle

\begin{abstract}
Studies of issues related to computability and computational complexity 
involve the use of a model of computation.
Pivotal to such a model are the computational processes considered.
Processes of this kind can be described using an imperative process 
algebra based on ACP (Algebra of Communicating Processes). 
In this paper, it is investigated whether the imperative process algebra 
concerned can play a role in the field of models of computation.
It is demonstrated that the process \linebreak[2] algebra is suitable to 
describe in a mathematically precise way models of computation 
corresponding to existing models based on sequential, \linebreak[2] 
asynchronous parallel, and synchronous parallel random access machines 
as well as time and work complexity measures for those models.
A probabilistic variant of the model based on sequential random access 
machines and complexity measures for it are also described.
\begin{classcode}
D.1.3, F.1.1, F.1.2, F.1.3. 
\end{classcode}
\end{abstract}

\section{Introduction}
\label{sect-intro}

A computational process is a process that solves a computational problem. 
A computational process is applied to a data environment that consists 
of data organized and accessible in a specific way.
Well-known examples of data environments are the tapes found in Turing 
machines and the memories found in random access machines.
The application of a computational process to a data environment yields
another data environment.
The data environment to which the process is applied represents an 
instance of the computational problem that is solved by the process and 
the data environment yielded by the application represents the 
solution of that instance.
A computational process is divided into simple steps, each of which 
depends on and has an impact on only a small portion of the data 
environment to which the process is applied.  

A basic assumption in this paper is that a model of computation is fully 
characterized by:
(a)~a set of possible computational processes,
(b)~for each possible computational process, a set of possible data 
environments, and
(c)~the effect of applying such processes to such environments.
The set of possible computational processes is usually given indirectly,
mostly by means of abstract machines that have a built-in program that
belongs to a set of possible programs which is such that the possible 
computational processes are exactly the processes that are produced by 
those machines when they execute their built-in program.
The abstract machines with their built-in programs emphasize the 
mechanical nature of the possible computational processes.
However, in this way, details of how possible computational processes 
are produced become part of the model of computation.
To the best of my knowledge, all definitions that have been given with 
respect to a model of computation and all results that have been proved 
with respect to a model of computation can do without reference to such
details. 

In~\cite{Mid21a}, an extension of ACP~\cite{BK84b} is presented whose 
additional features include assignment actions to change data in the 
course of a process, guarded commands to proceed at certain stages of 
a process in a way that depends on changing data, and data parameterized 
actions to communicate data between processes. 
The extension concerned is called \deACPet.
The term imperative process algebra was coined in~\cite{NP97a} for 
process algebras like \deACPet.
In~\cite{Mid21a}, it is discussed what qualities of \deACPet\ 
distinguish it from other imperative process algebras, how its 
distinguishing qualities are achieved, and what its relevance is to the 
verification of properties of processes carried out by contemporary 
computer-based systems.
Moreover, that paper goes into one of the application areas of \deACPet, 
namely the area of information-flow security analysis.

The current paper studies whether \deACPet\ can play a role in the field 
of models of computation.
The idea of this study originates from the experience that definitions 
of models of computation and results about them in the scientific 
literature tend to lack preciseness, in particular if it concerns models 
of parallel computation.
The study takes for granted the basic assumption about the 
characterization of models of computation mentioned above. 
Moreover, it focuses on models of computation that are intended for 
investigation of issues related to computability and computational 
complexity.
It does not consider models of computation geared to computation as it
takes place on concrete computers or computer networks of a certain 
kind.
Such models are left for follow-up studies.
Outcomes of this study are among other things mathematically precise 
definitions of models of computation corresponding to models based on 
sequential random access machines, asynchronous parallel random access 
machines, synchronous parallel random access machines, and a
probabilistic variant of sequential random access machines.

This paper is organized as follows.
First, a survey of the imperative process algebra \deACPet\ and its 
extension with recursion is given (Section~\ref{sect-deACPet}).
Next, it is explained in this process algebra what it means that a given 
process computes a given function (Section~\ref{sect-Computation}).
After that, a version of the sequential random access machine model of 
computation is described in the setting introduced in the previous two 
sections (Section~\ref{sect-RAM-Model}).
Following that, an asynchronous parallel random access machine model of 
computation and a synchronous parallel random access machine model of 
computation are described in that setting as well 
(Section~\ref{sect-APRAM-Model} and Section~\ref{sect-SPRAM-Model}, 
respectively).
Then, complexity measures for the models of computation presented in the
previous three sections are introduced 
(Section~\ref{sect-Measures}).
Thereafter, the question whether the presented synchronous parallel 
random access machine model of computation is a reasonable model of 
computation is treated (Section~\ref{sect-SPRAM-Model-more}).
Furthermore, a probabilistic variant of the presented random access 
machine model of computation is described 
(Section~\ref{sect-PrRAM-Model}).
Finally, some concluding remarks are made 
(Section~\ref{sect-conclusions}).

Section~\ref{sect-deACPet} is an abridged version of~\cite{Mid21a}.
Portions of Sections~2--4 of that paper have been copied verbatim or 
slightly modified.

\section{The Imperative Process Algebra \deACPet}
\label{sect-deACPet}

The imperative process algebra \deACPet\ is an extension of \ACPet, the 
version of \ACP\ with empty process and silent step constants that was 
first presented in~\cite[Section~5.3]{BW90}.
In this section, first a short survey of \ACPet\ is given and then 
\deACPet\ \linebreak[2] is introduced as an extension of \ACPet.
Moreover, recursion in the setting of \deACPet\ is treated and soundness 
and (semi-)completeness results for the axioms of \deACPet\ with 
recursion are presented.

\subsection{\ACP\ with Empty Process and Silent Step}
\label{subsect-ACPet}

In this section, a short survey of \ACPet\ is given.
A more comprehensive treatment of this process algebra can be found 
in~\cite{BW90}.

In \ACPet, it is assumed that a fixed but arbitrary finite set $\Act$ of 
\emph{basic actions}, with $\tau,\dead,\ep \not\in \Act$, and a fixed 
but arbitrary commutative and associative \emph{communication} function 
$\funct{\commf}
 {(\Act \Sunion \Set{\tau,\dead}) \Sx (\Act \Sunion \Set{\tau,\dead})}
 {(\Act \Sunion \Set{\tau,\dead})}$, 
such that $\commf(\tau,a) = \dead$ and $\commf(\dead,a) = \dead$
for all $a \in \Act \Sunion \Set{\tau,\dead}$, have been given.
Basic actions are taken as atomic processes.
The function $\commf$ is regarded to give the result of simultaneously
performing any two basic actions for which this is possible, and to be 
$\dead$ otherwise.
Henceforth, we write $\Actt$ for $\Act \Sunion \Set{\tau}$ and
$\Acttd$ for $\Act \Sunion \Set{\tau,\dead}$.

The algebraic theory \ACPet\ has one sort: the sort $\Proc$ of 
\emph{processes}.
This sort is made explicit to anticipate the need for many-sortedness 
later on. 
The algebraic theory \ACPet\ has the following constants and operators 
to build terms of sort~$\Proc$:
\begin{itemize}
\item
a \emph{basic action} constant $\const{a}{\Proc}$ for each 
$a \in \Act$;
\item
a \emph{silent step} constant $\const{\tau}{\Proc}$;
\item
a \emph{inaction} constant $\const{\dead}{\Proc}$;
\item
a \emph{empty process} constant $\const{\ep}{\Proc}$;
\item
a binary \emph{alternative composition} or \emph{choice} operator 
$\funct{\altc}{\Proc \Sx \Proc}{\Proc}$;
\item
a binary \emph{sequential composition} operator 
$\funct{\seqc}{\Proc \Sx \Proc}{\Proc}$;
\item
a binary \emph{parallel composition} or \emph{merge} operator 
$\funct{\parc}{\Proc \Sx \Proc}{\Proc}$;
\item
a binary \emph{left merge} operator 
$\funct{\leftm}{\Proc \Sx \Proc}{\Proc}$;
\item
a binary \emph{communication merge} operator 
$\funct{\commm}{\Proc \Sx \Proc}{\Proc}$;
\item
a unary \emph{encapsulation} operator 
$\funct{\encap{H}}{\Proc}{\Proc}$ for each $H \subseteq \Act$ and for 
$H = \Actt$; 
\item
a unary \emph{abstraction} operator 
$\funct{\abstr{I}}{\Proc}{\Proc}$ for each $I \subseteq \Act$.
\end{itemize}
It is assumed that there is a countably infinite set $\cX$ of variables 
of sort $\Proc$, which contains $x$, $y$ and $z$.
Terms are built as usual.
Infix notation is used for the binary operators.
The following precedence conventions are used to reduce the need for
parentheses: the operator ${} \seqc {}$ binds stronger than all other 
binary operators and the operator ${} \altc {}$ binds weaker than all 
other binary operators.

The constants $a$ ($a \in \Act$), $\tau$, $\ep$, and $\dead$ can be 
explained as follows:
\begin{itemize}
\item
$a$ denotes the process that first performs the observable action $a$ 
and then terminates successfully;
\item
$\tau$ denotes the process that first performs the unobservable action 
$\tau$ and then terminates successfully;
\item
$\ep$ denotes the process that terminates successfully without 
performing any action;
\item
$\dead$ denotes the process that cannot do anything, it cannot even 
terminate successfully.
\end{itemize}
Let $t$ and $t'$ be closed \ACPet\ terms denoting processes $p$ and 
$p'$, respectively. 
Then the operators $\altc$, $\seqc$, $\parc$, $\encap{H}$ 
($H \subseteq \Act$ or $H = \Actt$), and $\abstr{I}$ 
($I \subseteq \Act$) can be explained as follows:
\begin{itemize}
\item
$t \altc t'$ denotes the process that behaves as either $p$ or $p'$;
\item
$t \seqc t'$\, denotes the process that behaves as $p$ and $p'$ in 
sequence;
\item
$t \parc t'$ denotes the process that behaves as $p$ and $p'$ in 
parallel;
\item
$\encap{H}(t)$ denotes the process that behaves as $p$, except that 
actions from $H$ are blocked from being performed;
\item
$\abstr{I}(t)$\, denotes the process that behaves as $p$, except that 
actions from $I$ are turned into the unobservable action $\tau$.
\end{itemize}
The operators $\leftm$ and $\commm$ are of an auxiliary nature.
They make a finite axiomatization of \ACPet\ possible.

The axioms of \ACPet\ are presented in Table~\ref{axioms-ACPet}.
\begin{table}[!t]
\caption{Axioms of \ACPet}
\label{axioms-ACPet}
\begin{eqntbl}
\begin{axcol}
x \altc y = y \altc x                                & & \axiom{A1} \\
(x \altc y) \altc z = x \altc (y \altc z)            & & \axiom{A2} \\
x \altc x = x                                        & & \axiom{A3} \\
(x \altc y) \seqc z = x \seqc z \altc y \seqc z      & & \axiom{A4} \\
(x \seqc y) \seqc z = x \seqc (y \seqc z)            & & \axiom{A5} \\
x \altc \dead = x                                    & & \axiom{A6} \\
\dead \seqc x = \dead                                & & \axiom{A7} \\
x \seqc \ep = x                                      & & \axiom{A8} \\
\ep \seqc x = x                                      & & \axiom{A9} 
\eqnsep
x \parc y = x \leftm y \altc y \leftm x \altc x \commm y \altc
\encap{\Actt}\!(x) \seqc \encap{\Actt}\!(y)          & & \axiom{CM1E} \\
\ep \leftm x = \dead                                 & & \axiom{CM2E} \\
\alpha \seqc x \leftm y = \alpha \seqc (x \parc y)   & & \axiom{CM3}  \\
(x \altc y) \leftm z = x \leftm z \altc y \leftm z   & & \axiom{CM4}  \\
\ep \commm x = \dead                                 & & \axiom{CM5E} \\
x \commm \ep = \dead                                 & & \axiom{CM6E} \\
a \seqc x \commm b \seqc y = \commf(a,b) \seqc (x \parc y) 
                                                     & & \axiom{CM7}  \\
(x \altc y) \commm z = x \commm z \altc y \commm z   & & \axiom{CM8}  \\
x \commm (y \altc z) = x \commm y \altc x \commm z   & & \axiom{CM9}  
\eqnsep
\encap{H}(\ep) = \ep                                   & & \axiom{D0} \\
\encap{H}(\alpha) = \alpha      & \mif \alpha \notin H   & \axiom{D1} \\ 
\encap{H}(\alpha) = \dead       & \mif \alpha \in H      & \axiom{D2} \\
\encap{H}(x \altc y) = \encap{H}(x) \altc \encap{H}(y) & & \axiom{D3} \\
\encap{H}(x \seqc y) = \encap{H}(x) \seqc \encap{H}(y) & & \axiom{D4}
\eqnsep
\abstr{I}(\ep) = \ep                                   & & \axiom{T0} \\
\abstr{I}(\alpha) = \alpha      & \mif \alpha \notin I   & \axiom{T1} \\
\abstr{I}(\alpha) = \tau        & \mif \alpha \in I      & \axiom{T2} \\
\abstr{I}(x \altc y) = \abstr{I}(x) \altc \abstr{I}(y) & & \axiom{T3} \\
\abstr{I}(x \seqc y) = \abstr{I}(x) \seqc \abstr{I}(y) & & \axiom{T4} 
\eqnsep                                                                 
\alpha \seqc (\tau \seqc (x \altc y) \altc x) = \alpha \seqc (x \altc y)
                                                       & & \axiom{BE} 
\end{axcol}
\end{eqntbl}
\end{table}
In this table, $a$, $b$, and $\alpha$ stand for arbitrary members of 
$\Acttd$,\, $H$ stands for an arbitrary subset of $\Act$ 
or the \linebreak[2] set $\Actt$, and $I$ stands for an arbitrary subset 
of $\Act$.
So, CM3, CM7, D0--D4, T0--T4, and BE are actually axiom schemas.
In this paper, axiom schemas will usually be referred to as axioms.

The term $\encap{\Actt}\!(x) \seqc \encap{\Actt}\!(y)$ occurring in 
axiom CM1E is needed to handle successful termination in the presence of 
$\ep$:
it stands for the process that behaves the same as $\ep$ if both $x$ and 
$y$ stand for a process that has the option to behave the same as $\ep$ 
and it stands for the process that behaves the same as $\dead$ 
otherwise.
In~\cite[Section~5.3]{BW90}, the symbol $\surd$ is used 
instead of $\encap{\Actt}$.

Notice that the equation $\alpha \seqc \dead = \alpha$ would be 
derivable from the axioms of \ACPet\ if operators $\encap{H}$ where 
$H \neq \Actt$ and $\tau \in H$ were added to \ACPet.

The notation $\Altc{i=1}{n} t_i$, where $n \geq 1$, will be used for 
right-nested alternative compositions.
For each $n \in \Natpos$,%
\footnote
{We write $\Natpos$ for the set $\Set{n \in \Nat \where n \geq 1}$ of 
 positive natural numbers.} 
the term $\Altc{i = 1}{n} t_i$ is defined by induction on $n$ as 
follows:
$\Altc{i = 1}{1} t_i = t_1$ and 
$\Altc{i = 1}{n + 1} t_i = t_1 \altc \Altc{i = 1}{n} t_{i+1}$.
In addition, the convention will be used that 
$\Altc{i = 1}{0} t_i = \dead$.
Moreover, we write $\encap{a}$ and $\abstr{a}$, where $a \in \Act$, for 
$\encap{\Set{a}}$ and $\abstr{\Set{a}}$, respectively.

\subsection{Imperative \ACPet}
\label{subsect-deACPet}

In this section, \deACPet, imperative \ACPet, is introduced as an 
extension of \ACPet.
In~\cite{Mid21a}, the paper in which \deACPet\ was first presented, a
comprehensive treatment of this imperative process algebra can be found.
\deACPet\ extends \ACPet\ with features to communicate data between 
processes, to change data involved in a process in the course of the 
process, and to proceed at certain stages of a process in a way that 
depends on the changing data. 

In \deACPet, it is assumed that the following has been given with 
respect to data:
\begin{itemize}
\item
a many-sorted signature $\sign_\gD$ that includes:
\begin{itemize}
\item
a sort $\Data$ of \emph{data} and
a sort $\Bool$ of \emph{bits};
\item
constants of sort $\Data$ and/or operators with result sort $\Data$;
\item
constants $\zero$ and $\one$ of sort $\Bool$ and
operators with result sort $\Bool$;
\end{itemize}
\item
a minimal algebra $\gD$ of the signature $\sign_\gD$ in which 
the carrier of sort $\Bool$ has cardinality $2$ and 
the equation $\zero = \one$ does not hold.
\end{itemize}
In \deACPet, it is moreover assumed that a finite or countably infinite 
set $\FlexVar$ of \emph{flexible variables} has been given.
A flexible variable is a variable whose value may change in the course 
of a process.%
\footnote
{The term flexible variable is used for this kind of variables in 
 e.g.~\cite{Sch97a,Lam94a}.} 

We write $\DataVal$ for the set of all closed terms over the signature
$\sign_\gD$ that are of sort~$\Data$.

A \emph{flexible variable valuation} is a function from $\FlexVar$ to 
$\DataVal$. 
Flexible variable valuations are intended to provide the data values 
--- which are members of $\gD$'s carrier of sort $\Data$ --- assigned to 
flexible variables when an \deACPet\ term of sort $\Data$ is evaluated.
To fit better in an algebraic setting, they provide closed terms from 
$\DataVal$ that denote those data values instead.

Because $\gD$ is a minimal algebra, for each sort $S$ that is included 
in $\sign_\gD$, each member of $\gD$'s carrier of sort $S$ can be 
represented by a closed term over $\sign_\gD$ that is of sort $S$. 

In the rest of this paper, for each sort $S$ that is included in 
$\sign_\gD$, let $\nm{ct}_S$ be a function from $\gD$'s carrier of sort 
$S$ to the set of all closed terms over $\sign_\gD$ that are of sort~$S$ 
such that, for each member $d$ of $\gD$'s carrier of sort $S$, the term 
$\nm{ct}_S(d)$ represents $d$. 
\label{def-ct}
We write $d$, where $d$ is a member of $\gD$'s carrier of sort $S$, for 
$\nm{ct}_S(d)$ if it is clear from the context that $d$ stands for a 
closed term of sort $S$ representing $d$.

Flexible variable valuations are used in 
Sections~\ref{sect-RAM-Model}--\ref{sect-SPRAM-Model}
and~\ref{sect-PrRAM-Model} to represent the data enviroments referred to 
in Section~\ref{sect-intro}.

Let $V \subseteq \FlexVar$.
Then a \emph{$V$-indexed data environment} is a function from $V$ to  
$\gD$'s carrier of sort $\Data$. 
Let $\mu$ be a $V$-indexed data environment and $\rho$ be a flexible 
variable valuation.
Then $\rho$ \emph{represents} $\mu$ if 
$\rho(v) = \nm{ct}_\Data(\mu(v))$ for all $v \in V$.

Below, the sorts, constants and operators of \deACPet\ are introduced.
The operators of \deACPet\ include two variable-binding operators.
\sloppy
The formation rules for \mbox{\deACPet}\ terms are the usual ones for 
the many-sorted case (see e.g.~\cite{ST99a,Wir90a}) and in addition the 
following rule:
\begin{itemize}
\item
if $O$ is a variable-binding operator 
$\funct{O}{S_1 \Sx \ldots \Sx S_n}{S}$ that binds a variable of sort~$S'$,
$t_1,\ldots,t_n$~are terms of sorts $S_1,\ldots,S_n$, respectively, and 
$X$ is a variable of sort $S'$, then $O X (t_1,\ldots,t_n)$ is a term of 
sort $S$.
\end{itemize}
An extensive formal treatment of the phenomenon of variable-binding 
operators can be found in~\cite{PS95a}.

\deACPet\ has the following sorts: 
the sorts included in $\sign_\gD$,
the sort $\Cond$ of \emph{conditions}, and
the sort $\Proc$ of \emph{processes}.

For each sort $S$ included in $\sign_\gD$ other than $\Data$, 
\deACPet\ has only the constants and operators included in $\sign_\gD$ 
to build terms of sort $S$.

\deACPet\ has, in addition to the constants and operators included in 
$\sign_\gD$ to build terms of sorts $\Data$, the following constants to 
build terms of sort $\Data$:
\begin{itemize}
\item
for each $v \in \FlexVar$, the \emph{flexible variable} constant 
$\const{v}{\Data}$.
\end{itemize}
We write $\DataTerm$ for the set of all closed \deACPet\ terms of sort 
$\Data$.

\deACPet\ has the following constants and operators to build terms of 
sort~$\Cond$:
\begin{itemize}
\item
a binary \emph{equality} operator
$\funct{\Leq}{\Bool \Sx \Bool}{\Cond}$;
\item
a binary \emph{equality} operator
$\funct{\Leq}{\Data \Sx \Data}{\Cond}$;%
\footnote
{The overloading of $=$ can be trivially resolved if $\sign_\gD$ is
 without overloaded symbols.}
\item
a \emph{truth} constant $\const{\True}{\Cond}$;
\item
a \emph{falsity} constant $\const{\False}{\Cond}$;
\item
a unary \emph{negation} operator $\funct{\Lnot}{\Cond}{\Cond}$;
\item
a binary \emph{conjunction} operator 
$\funct{\Land}{\Cond \Sx \Cond}{\Cond}$;
\item
a binary \emph{disjunction} operator 
$\funct{\Lor}{\Cond \Sx \Cond}{\Cond}$;
\item
a binary \emph{implication} operator 
$\funct{\Limpl}{\Cond \Sx \Cond}{\Cond}$;
\item
a unary variable-binding \emph{universal quantification} operator 
$\funct{\forall}{\Cond}{\Cond}$ that binds a variable of sort $\Data$; 
\item
a unary variable-binding \emph{existential quantification} operator 
$\funct{\exists}{\Cond}{\Cond}$ that binds a variable of sort $\Data$. 
\end{itemize}
We write $\CondTerm$ for the set of all closed \deACPet\ terms of sort 
$\Cond$.

\deACPet\ has, in addition to the constants and operators of \ACPet, 
the following operators to build terms of sort $\Proc$:
\begin{itemize}
\item
an $n$-ary \emph{data parameterized action} operator
$\funct{a}{\Data ^n}{\Proc}$ for each $a \in \Act$, for each 
$n \in \Nat$;
\item
a unary \emph{assignment action} operator
$\funct{\assop{v}\,}{\Data}{\Proc}$ for each $v \in \FlexVar$;
\item
a binary \emph{guarded command} operator 
$\funct{\gc\,}{\Cond \Sx \Proc}{\Proc}$;
\item
a unary \emph{evaluation} operator 
$\funct{\eval{\rho}}{\Proc}{\Proc}$ for each $\rho \in \FVarVal$.
\end{itemize}
We write $\ProcTerm$ for the set of all closed \deACPet\ terms of sort 
$\Proc$.

It is assumed that there are countably infinite sets of variables of 
sort $\Data$ and $\Cond$ and that the sets of variables of sort $\Data$, 
$\Cond$, and $\Proc$ are mutually disjoint and disjoint from $\FlexVar$.

The same notational conventions are used as before.
Infix notation is also used for the additional binary operators.
Moreover, the notation $\ass{v}{e}$, where $v \in \FlexVar$ and $e$ is a 
\deACPet\ term of sort $\Data$, is used for the term $\assop{v}(e)$.

Each term from $\CondTerm$ can be taken as a formula of a first-order 
language with equality of $\gD$ by taking the flexible variable
constants as additional variables of sort $\Data$.
The flexible variable constants are implicitly taken as additional 
variables of sort $\Data$ wherever the context asks for a formula.
In this way, each term from $\CondTerm$ can be interpreted in $\gD$ as a
formula.

The notation $\phi \Liff \psi$, where $\phi$ and $\psi$ are 
\deACPet\ terms of sort $\Cond$, is used for the term
$(\phi \Limpl \psi) \Land (\psi \Limpl \phi)$.
The axioms of \deACPet\ (given below) include an equation $\phi = \psi$ 
for each two terms $\phi$ and $\psi$ from $\CondTerm$ for which the 
formula $\phi \Liff \psi$ holds in $\gD$.

Let 
$a$ be a basic action from $\Act$, 
$e_1$, \ldots, $e_n$, and $e$ be terms from $\DataTerm$,  
$\phi$ be a term from $\CondTerm$, and 
$t$ be a term from $\ProcTerm$ denoting a process $p$.
Then the additional operators to build terms of sort $\Proc$ can be 
explained as follows:
\begin{itemize}
\item
the term $a(e_1,\ldots,e_n)$ denotes the process that first performs the 
data parameterized action $a(e_1,\ldots,e_n)$ and then terminates 
successfully;
\item
the term $\ass{v}{e}$ denotes the process that first performs the 
assignment action $\ass{v}{e}$, whose intended effect is the assignment 
of the result of evaluating $e$ to flexible variable $v$, and then 
terminates successfully; 
\item
the term $\phi \gc t$ denotes the process that behaves as $p$ if 
condition $\phi$ holds and as $\dead$ otherwise;
\item
the term $\eval{\rho}(t)$ denotes the process that behaves as $p$, 
except that each subterm of $t$ that belongs to $\DataTerm$ is evaluated 
using flexible variable valuation $\rho$ updated according to the 
assignment actions that have taken place at the point where the subterm 
is encountered.
\end{itemize}
Evaluation operators are a variant of state operators 
(see e.g.~\cite{BB88}).

The following closed \deACPet\ term is reminiscent of a program that 
computes the difference between two integers by subtracting the smaller 
one from the larger one ($i, j, d \in \FlexVar$):
\begin{ldispl}
\ass{d}{i} \seqc  
((d \geq j = \one)  \gc \ass{d}{d - j} \altc 
 (d \geq j = \zero) \gc \ass{d}{j - d})\;.%
\footnotemark
\end{ldispl}%
\footnotetext
{Here and in the next example, the carrier of $\Data$ is assumed to be 
 the set of all integers. Moreover, the usual integer constants,
 operators  on integers, and predicates on integers are assumed (where
 operators with result sort $\Bool$ serve as predicates).}%
That is, the final value of $d$ is the absolute value of the result of
subtracting the initial value of $i$ from the initial value of $j$.
Let $\rho$ be an flexible variable valuation such that $\rho(i) = 11$ 
and $\rho(j) = 3$.
Then the following equation can be derived from the axioms of \deACPet\
given below:
\begin{ldispl}
\eval{\rho}
(\ass{d}{i} \seqc  
 ((d \geq j = \one)  \gc \ass{d}{d - j} \altc
  (d \geq j = \zero) \gc \ass{d}{j - d})) \\
\; {} =
\ass{d}{11} \seqc \ass{d}{8}\;. 
\end{ldispl}%
This equation shows that in the case where the initial values of $i$ 
and $j$ are $11$ and $3$ the final value of $d$ is $8$, which is the 
absolute value of the result of subtracting $11$ from $3$.

A flexible variable valuation $\rho$ can be extended homomorphically 
from flexible variables to \deACPet\ terms of sort $\Data$ and \deACPet\ 
terms of sort $\Cond$.
Below, these extensions are denoted by $\rho$ as well.
Moreover, we write $\rho\mapupd{e}{v}$ for the flexible variable 
valuation $\rho'$ defined by $\rho'(v') = \rho(v')$ if $v' \neq v$ and 
$\rho'(v) = e$.

The subsets $\AProcPAR$, $\AProcASS$, and $\AProcTerm$ of $\ProcTerm$ 
referred to below are defined as follows:
\begin{ldispl}
\begin{aeqns}
\AProcPAR  & = & {} \Union_{n \in \Natpos}
\Set{a(e_1,\dots,e_n) \where
     a \in \Act \Land e_1,\dots,e_n \in \DataTerm}\;, \\
\AProcASS & = & 
\Set{\ass{v}{e} \where v \in \FlexVar \Land e \in \DataTerm}\;, \\
\AProcTerm & = & 
\Act \Sunion \AProcPAR \Sunion \AProcASS\;.
\end{aeqns}
\end{ldispl}%
The elements of $\AProcTerm$ are the terms from $\ProcTerm$ that denote 
the processes that are considered to be atomic.
Henceforth, we write $\AProcTermt$ for $\AProcTerm \Sunion \Set{\tau}$,
$\AProcTermd$ for $\AProcTerm \Sunion \Set{\dead}$, and $\AProcTermtd$ 
for $\AProcTerm \Sunion \Set{\tau,\dead}$.

The axioms of \deACPet\ are the axioms presented in 
Tables~\ref{axioms-ACPet} and~\ref{axioms-deACPet},
\begin{table}[!t]
\caption{Additional axioms of \deACPet}
\label{axioms-deACPet}
\begin{eqntbl}
\begin{axcol}
e = e'         & \mif \Sat{\gD}{\fol{e = e'}}          & \axiom{IMP1} \\
\phi = \psi    & \mif \Sat{\gD}{\fol{\phi \Liff \psi}} & \axiom{IMP2} 
\eqnsep
\True \gc x = x                                      & & \axiom{GC1}  \\
\False \gc x = \dead                                 & & \axiom{GC2}  \\
\phi \gc \dead = \dead                               & & \axiom{GC3}  \\
\phi \gc (x \altc y) = \phi \gc x \altc \phi \gc y   & & \axiom{GC4}  \\
\phi \gc x \seqc y = (\phi \gc x) \seqc y            & & \axiom{GC5}  \\
\phi \gc (\psi \gc x) = (\phi \Land \psi) \gc x      & & \axiom{GC6}  \\
(\phi \Lor \psi) \gc x = \phi \gc x \altc \psi \gc x & & \axiom{GC7}  \\
(\phi \gc x) \leftm y = \phi \gc (x \leftm y)        & & \axiom{GC8}  \\
(\phi \gc x) \commm y = \phi \gc (x \commm y)        & & \axiom{GC9}  \\
x \commm (\phi \gc y) = \phi \gc (x \commm y)        & & \axiom{GC10} \\
\encap{H}(\phi \gc x) = \phi \gc \encap{H}(x)        & & \axiom{GC11} \\
\abstr{I}(\phi \gc x) = \phi \gc \abstr{I}(x)        & & \axiom{GC12} 
\eqnsep
\eval{\rho}(\ep) = \ep                               & & \axiom{V0}   \\
\eval{\rho}(\alpha \seqc x) = \alpha \seqc \eval{\rho}(x)
      & \mif \alpha \notin \AProcPAR \Sunion \AProcASS & \axiom{V1}   \\
\eval{\rho}(a(e_1,\ldots,e_n) \seqc x) = 
a(\rho(e_1),\ldots,\rho(e_n)) \seqc \eval{\rho}(x)
                                                     & & \axiom{V2}   \\
\eval{\rho}(\ass{v}{e} \seqc x) = 
{\ass{v}{\rho(e)} \seqc \eval{\rho\mapupd{\rho(e)}{v}}(x)} 
                                                     & & \axiom{V3}   \\
\eval{\rho}(x \altc y) = \eval{\rho}(x) \altc \eval{\rho}(y)
                                                     & & \axiom{V4}   \\
\eval{\rho}(\phi \gc y) = \rho(\phi) \gc \eval{\rho}(x)
                                                     & & \axiom{V5}   
\eqnsep
a(e_1,\ldots,e_n) \seqc x \commm b(e'_1,\ldots,e'_n) \seqc y = 
 {} \\ \quad
(e_1 = e'_1 \Land \ldots \Land e_n = e'_n) \gc c(e_1,\ldots,e_n) \seqc
(x \parc y)                    & \mif \commf(a,b) = c & \axiom{CM7Da} \\
a(e_1,\ldots,e_n) \seqc x \commm b(e'_1,\ldots,e'_m) \seqc y = \dead
  & \mif \commf(a,b) = \dead \;\mathrm{or}\; n \neq m & \axiom{CM7Db} \\
a(e_1,\ldots,e_n) \seqc x \commm \alpha \seqc y = \dead 
 & \mif \alpha \notin \AProcPAR & \axiom{CM7Dc} \\
\alpha \seqc x \commm a(e_1,\ldots,e_n) \seqc y = \dead 
 & \mif \alpha \notin \AProcPAR & \axiom{CM7Dd} \\
\ass{v}{e} \seqc x \commm \alpha \seqc y = \dead    & & \axiom{CM7De} \\
\alpha \seqc x \commm \ass{v}{e} \seqc y = \dead    & & \axiom{CM7Df}   
\eqnsep
\alpha \seqc (\phi \gc \tau \seqc (x \altc y) \altc \phi \gc x) = 
\alpha \seqc (\phi \gc (x \altc y))                 & & \axiom{BED}
\end{axcol}
\end{eqntbl}
\end{table}
where
$\alpha$ stands for an arbitrary term from $\AProcTermtd$,\, 
\pagebreak[2]
$H$ stands for an arbitrary subset of $\AProcTerm$ or the set 
$\AProcTermt$, \pagebreak[2]
$I$ stands for an arbitrary subset of $\AProcTerm$,\, 
$e$, $e_1,e_2,\ldots$\ and $e'$, $e'_1,e'_2,\ldots$\ stand for 
arbitrary terms from $\DataTerm$,\, 
$\phi$ and $\psi$ stand for arbitrary terms from $\CondTerm$,
$v$ stands for an arbitrary flexible variable from $\FlexVar$, and
$\rho$ stands for an arbitrary flexible variable valuation from 
$\FVarVal$.
Moreover,\, $a$, $b$, and $c$ stand for arbitrary members of $\Acttd$ in 
Table~\ref{axioms-ACPet} and for arbitrary members of $\Act$ in 
Table~\ref{axioms-deACPet}.

\subsection{\deACPet\ with Recursion}
\label{subsect-deACPetr}

In this section, recursion in the setting of \deACPet\ is treated.
A closed \deACPet\ term of sort $\Proc$ denotes a process with a finite 
upper bound to the number of actions that it can perform. 
Recursion allows the description of processes without a finite upper 
bound to the number of actions that it can perform.

A \emph{recursive specification} over \deACPet\ is a set 
$\Set{X_i = t_i \where i \in I}$, where $I$ is a finite set, 
each $X_i$ is a variable from $\cX$, each $t_i$ is a \deACPet\ term of 
sort $\Proc$ in which only variables from $\Set{X_i \where i \in I}$ 
occur, and $X_i \neq X_j$ for all $i,j \in I$ with $i \neq j$. 
We write $\vars(E)$, where $E$ is a recursive specification over 
\deACPet, for the set of all variables that occur in $E$.
Let $E$ be a recursive specification and let $X \in \vars(E)$.
Then there exists a unique equation in $E$ whose left-hand side is $X$.
This equation is called the \emph{recursion equation for $X$ in $E$}.

Below, guarded linear recursive specifications over \deACPet\ are 
introduced.
The set $\LT$ of \emph{linear \deACPet\ terms} is inductively defined by 
the following rules:
\begin{enumerate}
\item
$\dead \in \LT$;
\item 
if $\phi \in \CondTerm$, then $\phi \gc \ep \in \LT$;
\item
if $\phi \in \CondTerm$, $\alpha \in \AProcTermt$, and $X \in \cX$, then 
$\phi \gc \alpha \seqc X \in \LT$;
\item
if $t,t' \in \LT \Sdiff \Set{\dead}$, then $t \altc t' \in \LT$.
\end{enumerate}
Let $X$ be a variable from $\cX$ and
let $t$ be an \deACPet\ term in which $X$ occurs. 
Then an occurrence of $X$ in $t$ is \emph{guarded} if $t$ has a subterm 
of the form $\alpha \seqc t'$ where $\alpha \in \AProcTerm$ and $t'$ 
contains this occurrence of $X$.
Notice that an occurrence of a variable in a linear \deACPet\ term may 
be not guarded. 

A \emph{guarded linear recursive specification} over \deACPet\ is a 
recursive specification $\Set{X_i = t_i \where i \in I}$ over \deACPet\ 
where each $t_i$ is a linear \deACPet\ term, and there does not exist an 
infinite sequence $i_0\;i_1\;\ldots\,$ over $I$ such that, for each 
$k \in \Nat$, there is an occurrence of $X_{i_{k+1}}$ in $t_{i_k}$ 
that is not guarded.

A solution of a guarded linear recursive specification $E$ over 
\deACPet\ in some model of \deACPet\ is a set 
$\Set{p_X \where X \in \vars(E)}$ of elements of the carrier of sort
$\Proc$ in that model such that each equation in $E$ holds if, for all 
$X \in \vars(E)$, \linebreak[2] $X$ is assigned $p_X$. 
A guarded linear recursive specification has a unique solution under the
equivalence defined in~\cite{Mid21a} for \deACPet\ extended with guarded 
linear recursion.
If $\Set{p_X \where X \in \vars(E)}$ is the unique solution of a guarded 
linear recursive specification $E$, then, for 
each $X \in \vars(E)$, $p_X$ is called the \emph{$X$-component} of the 
unique solution of $E$. 

\deACPet\ is extended with guarded linear recursion by adding constants 
for solutions of guarded linear recursive specifications over \deACPet\ 
and axioms concerning these additional constants.
For each guarded linear recursive specification $E$ over \deACPet\ and 
each $X \in \vars(E)$, a constant $\rec{X}{E}$ of sort $\Proc$, that 
stands for the $X$-component of the unique solution of $E$, is added to 
the constants of \deACPet.
The equation RDP and the conditional equation RSP given in 
Table~\ref{axioms-REC} are added to the axioms of \deACPet.
\begin{table}[!t]
\caption{Axioms for guarded linear recursion}
\label{axioms-REC}
\begin{eqntbl}
\begin{axcol}
\rec{X}{E} = \rec{t}{E} & \mif X \!= t \;\in\, E & \axiom{RDP} \\
E \Limpl X = \rec{X}{E} & \mif X \in \vars(E)    & \axiom{RSP} 
\end{axcol}
\end{eqntbl}
\end{table}
In this table, $X$ stands for an arbitrary variable from $\cX$, 
$t$ stands for an arbitrary \deACPet\ term of sort $\Proc$,\, 
$E$ stands for an arbitrary guarded linear recursive specification over 
\deACPet, and 
the notation $\rec{t}{E}$ is used for $t$ with, for all 
$X \in \vars(E)$, all oc\-cur\-rences of $X$ in $t$ replaced by 
$\rec{X}{E}$.
Side conditions restrict what $X$, $t$ and $E$ stand for.

We write \deACPetr\ for the resulting theory.
Furthermore, we write $\ProcTermr$ for the set of all closed 
$\deACPetr$ terms of sort $\Proc$.

RDP and RSP together postulate that guarded linear recursive 
specifications over \deACPet\ have unique solutions.

Because RSP introduces conditional equations in \deACPetr, it is 
understood that conditional equational logic is used in deriving 
equations from the axioms of \deACPetr.
A complete inference system for conditional equational logic can for
example be found in~\cite{BW90,Gog21a}.

The following closed \deACPetr\ term is reminiscent of a program that 
computes by repeated subtraction the quotient and remainder of dividing a
non-negative integer by a positive integer ($i, j, q, r \in \FlexVar$):
\begin{ldispl}
\ass{q}{0} \seqc \ass{r}{i} \seqc \rec{Q}{E}\;,  
\end{ldispl}%
where $E$ is the guarded linear recursive specification that consists of 
the following two equations ($Q,R \in \cX$):
\begin{ldispl}
Q = (r \geq j = \one)  \gc \ass{q}{q + 1} \seqc R \altc 
    (r \geq j = \zero) \gc \ep\;, \\
R = \True \gc \ass{r}{r - j} \seqc Q\;.
\end{ldispl}%
Let $\rho$ be an flexible variable valuation such that $\rho(i) = 11$ 
and $\rho(j) = 3$.
Then the following equation can be derived from the axioms of \deACPetr:
\begin{ldispl}
\eval{\rho}
(\ass{q}{0} \seqc \ass{r}{i} \seqc \rec{Q}{E}) \\
\; {} =
\ass{q}{0} \seqc \ass{r}{11} \seqc \ass{q}{1} \seqc \ass{r}{8} \seqc 
\ass{q}{2} \seqc \ass{r}{5} \seqc \ass{q}{3} \seqc \ass{r}{2}\;. 
\end{ldispl}%
This equation shows that in the case where the initial values of $i$ 
and $j$ are $11$ and $3$ the final values of $q$ and $r$ are $3$ and 
$2$, which are the quotient and remainder of dividing $11$ by $3$.

In~\cite{Mid21a}, an equational axiom schema CFAR (Cluster Fair 
Abstraction Rule) is added to \deACPetr.
CFAR expresses that every cluster of $\tau$ actions will be exited 
sooner or later.
This is a fairness assumption made in the verification of many 
properties concerning the external behaviour of systems.
We will write \deACPetrf\ for the theory \deACPetr\ extended with CFAR.

We write $T \Ent t = t'$, where $T$ is \deACPetr\ or \deACPetrf, to 
indicate that the equation $t = t'$ is derivable from the axioms of $T$ 
using a complete inference system for conditional equational logic.

\subsection{Soundness and Completeness Results}
\label{subsect-sound-compl}

In~\cite{Mid21a}, a structural operational semantics of \deACPetr\ is 
presented and an equivalence relation $\rbbisim$ on $\ProcTermr$ based 
on this structural operational semantics is defined.
This equivalence relation reflects the idea that two processes are 
equivalent if they can simulate each other insofar as their observable 
potentials to make transitions by performing actions and to terminate 
successfully are concerned, taking into account the assigments of data 
values to flexible variables under which the potentials are available.

In this section, soundness and (semi-)completeness results for the 
axioms of \deACPetrf\ with respect to $\rbbisim$ are presented.
The proofs can be found in~\cite{Mid21a}.

The axiom system of \deACPetrf\ is sound with respect to ${\rbbisim}$ 
for equations between terms from $\ProcTermr$.
\begin{theorem}[Soundness]
\label{theorem-soundness-ACPet}
For all terms $t,t' \in \ProcTermr$, $t = t'$ is derivable from the 
axioms of \deACPetrf\ only if $t \rbbisim t'$.
\end{theorem}

The axiom system of \deACPetrf\ is incomplete with respect to $\rbbisim$ 
for equations between terms from $\ProcTermr$ and there is no 
straightforward way to rectify this.
Below two semi-completeness results are presented.
The next two lemmas are used in the proofs of those results.

A term $t \in \ProcTermr$ is called \emph{abstraction-free} if no 
abstraction operator occurs in~$t$.
A term $t \in \ProcTermr$ is called \emph{bool-conditional} if,
for each  $\phi \in \CondTerm$ that occurs in~$t$, 
$\Sat{\gD}{\fol{\phi \Liff \True}}$ or 
$\Sat{\gD}{\fol{\phi \Liff \False}}$.

\begin{lemma}
\label{lemma-glr-abstr-free}
For all abstraction-free $t \in \ProcTermr$, there exists a guarded
lin\-ear re\-cursive specification $E$ and $X \in \vars(E)$ such that
$\deACPetr \Ent \linebreak[2] t = \rec{X}{E}$.
\end{lemma}
\begin{lemma}
\label{lemma-glr-bool-cond}
\sloppy
For all bool-conditional $t \in \ProcTermr$, there exists a guarded
linear recursive specification $E$ and $X \in \vars(E)$ such that
$\deACPetrf \Ent \mbox{$t = \rec{X}{E}$}$.
\end{lemma}

The following two theorems are the semi-completeness results referred to 
above.
\begin{theorem}[Semi-completeness I]
\label{theorem-completeness-ACPet-1}
\sloppy
For all abstraction-free $t,t' \in \ProcTermr$, 
$\deACPetr \Ent \mbox{$t = t'$}$ if $t \rbbisim t'$.
\end{theorem}
\begin{theorem}[Semi-completeness II]
\label{theorem-completeness-ACPet-2}
For all bool-conditional $t,t' \in \ProcTermr$, 
$\deACPetrf \Ent \mbox{$t = t'$}$ if $t \rbbisim t'$.
\end{theorem}

\subsection{Results about the Evaluation Operators}
\label{subsect-eval-op}

For a better understanding of the evaluation operators, some results 
about these rather unfamiliar operators are given in this section.

The following lemma tells us that a closed term of the form 
$\eval{\rho}(t)$ equals a bool-conditional closed term.
\begin{lemma}
\label{lemma-eval-bool-cond}
\sloppy
For all $t \in \ProcTermr$, for all $\rho \in \FVarVal$, there exists a 
bool-conditional $t' \in \ProcTermr$ such that 
$\deACPetrf \Ent \eval{\rho}(t) = t'$.
\end{lemma}
\begin{proof}
This is straightforwardly proved by induction on the length of $t$, 
case distinction on the structure of $t$, and in the case of the 
constants for solutions of guarded linear recursive specifications 
additionally by induction on the structure of the right-hand side of
a recursion equation.
\qed
\end{proof}

The following theorem is a soundness and completeness result for closed
terms of the form $\eval{\rho}(t)$.
\begin{theorem}
\label{theorem-completeness-ACPet}
For all $t,t' \in \ProcTermr$, for all $\rho \in \FVarVal$, 
$\deACPetrf \Ent \eval{\rho}(t) = \eval{\rho}(t')$ iff 
$\eval{\rho}(t) \rbbisim \eval{\rho}(t')$.
\end{theorem}
\begin{proof}
This follows immediately from Theorem~\ref{theorem-soundness-ACPet},
Theorem~\ref{theorem-completeness-ACPet-2}, and 
Lemma~\ref{lemma-eval-bool-cond}.
\qed
\end{proof}

Below, an elimination theorem for closed terms of the form 
$\eval{\rho}(t)$ is presented.
In preparation, the subsets $\BT$ and $\BTcf$ of $\ProcTerm$ are 
introduced. 

The set $\BT$ of \emph{basic \deACPet\ terms} is inductively defined by 
the following rules:
\begin{enumerate}
\item
$\dead \in \BT$;
\item 
if $\phi \in \CondTerm$, then $\phi \gc \ep \in \BT$;
\item
if $\phi \in \CondTerm$, $\alpha \in \AProcTermt$, and $t \in \BT$, then 
$\phi \gc \alpha \seqc t \in \BT$;
\item
if $t,t' \in \BT \Sdiff \Set{\dead}$, then $t \altc t' \in \BT$.
\end{enumerate}
\begin{lemma}
\label{lemma-elim-1}
\sloppy
For all bool-conditional $t \in \ProcTerm$, there exists a 
bool-conditional $t' \in \BT$ such that $\deACPet \Ent t = t'$.
\end{lemma}
\begin{proof}
This is straightforwardly proved by induction on the length of $t$ and 
case distinction on the structure of $t$.
\qed
\end{proof}

The set $\BTcf$ of \emph{condition-free basic \deACPet\ terms} is 
inductively defined by the following rules:
\begin{enumerate}
\item
$\dead \in \BTcf$;
\item 
$\ep \in \BTcf$;
\item
if $\alpha \in \AProcTermt$, and $t \in \BTcf$, then 
$\alpha \seqc t \in \BTcf$;
\item
if $t,t' \in \BTcf \Sdiff \Set{\dead}$, then $t \altc t' \in \BTcf$.
\end{enumerate}
\begin{lemma}
\label{lemma-elim-2}
\sloppy
For all bool-conditional $t \in \BT$, there exists a $t' \in \BTcf$ such 
that $\deACPet \Ent t = t'$.
\end{lemma}
\begin{proof}
This is easily proved by induction on the structure of $t$.
\qed
\end{proof}

A term $t \in \ProcTermr$ is called a \emph{finite process term} if 
there exists a term $t' \in \ProcTerm$ such that 
$\deACPetrf \Ent t = t'$.

The following theorem tells us that a finite process term of the form 
$\eval{\rho}(t)$ equals a condition-free basic term.
\begin{theorem}
\label{theorem-elim}
For all $t \in \ProcTermr$ and $\rho \in \FVarVal$ for which 
$\eval{\rho}(t)$ is a finite process term, there exists a 
$t' \in \BTcf$ such that $\deACPetrf \Ent \eval{\rho}(t) = t'$.
\end{theorem}
\begin{proof}
This follows immediately from Lemmas~\ref{lemma-eval-bool-cond}, 
\ref{lemma-elim-1}, and~\ref{lemma-elim-2}.
\qed
\end{proof}
The terms from $\BTcf$ are reminiscent of computation trees.
In Section~\ref{sect-Computation}, use is made of the fact that
each finite process term of the form $\eval{\rho}(t)$ equals such a 
term.

Not every term from $\BTcf$ corresponds to a computation tree of which 
each path represents a computation that eventually halts, not even when 
it concerns a computation tree with a single path.

A term $t \in \ProcTermr$ is called a \emph{terminating process term} if 
there exists a term $t' \in \BTcf$ such that $\deACPetrf \Ent t = t'$ 
and $t'$ can be formed by applying only the formation rules~2, 3, and~4 
of $\BTcf$.

\subsection{Extensions}
\label{subsect-extensions}

In this section, two extensions of \deACPet\ are treated, namely an
extension with projection and an extension with action renaming.
It is not unusual to come across these extensions in applications of 
\ACP-style process algebras.
The first extension is treated here because projections can be used 
to determine the maximum number of actions that a finite process can
perform.
The second extension is treated here because action renaming enables 
to easily define the synchronous variant of the parallel composition 
operator of \deACPet\ needed later in this paper.

Let $T$ be either \deACPet\ or one of its extensions introduced before.
$T$ can be extended with projection by adding, for each $n \in \Nat$, 
a unary \emph{projection} operator $\funct{\proj{n}}{\Proc}{\Proc}$ 
to the operators of $T$ and adding the axioms given in 
Table~\ref{axioms-proj} to the axioms of $T$.
\begin{table}[!t]
\caption{Axioms for the projection operators}
\label{axioms-proj}
\begin{eqntbl}
\begin{axcol}
\proj{n}(\ep) = \ep                                     & \axiom{PR1} \\
\proj{0}(\alpha \seqc x) = \ep                          & \axiom{PR2} \\
\proj{n+1}(\alpha \seqc x) = \alpha \seqc \proj{n}(x)   & \axiom{PR3} \\
\proj{n}(x \altc y) = \proj{n}(x) \altc \proj{n}(y)     & \axiom{PR4} \\
\proj{n}(\phi \gc x) = \phi \gc \proj{n}(x)             & \axiom{PR5} \\
\proj{n}(\tau \seqc x) = \tau \seqc \proj{n}(x)         & \axiom{PR6}
\end{axcol}
\end{eqntbl}
\end{table}
In this table, 
$n$ stands for an arbitrary natural number,
$\alpha$ stands for an arbitrary term from $\AProcTermd$, and
$\phi$ stands for an arbitrary term from $\CondTerm$.

Let $t$ is a closed term of the extended theory.
Then the projection operator $\proj{n}$ can be explained as follows: 
$\proj{n}(t)$ denotes the process that behaves the same as the 
process denoted by $t$ except that it terminates successfully after $n$ 
actions have been performed.

Let $T$ be either \deACPet\ or one of its extensions introduced before.
Then we will write $T$+\PR\ for $T$ extended with the projection 
operators $\proj{n}$ and the axioms PR1--PR6 from 
Table~\ref{axioms-proj}.

Let $T$ be either \deACPet\ or one of its extensions introduced before.
$T$ can be extended with action renaming by adding, 
for each function $\funct{f}{\AProcTerm}{\AProcTerm}$ such that 
$f(\alpha) = \alpha$ for all $\alpha \in \AProcASS$, 
a unary \emph{action renaming} operator $\funct{\rnm{f}}{\Proc}{\Proc}$ 
to the operators of $T$ and adding the axioms given in 
Table~\ref{axioms-rnm} to the axioms of $T$.
\begin{table}[!t]
\caption{Axioms for the action renaming operators}
\label{axioms-rnm}
\begin{eqntbl}
\begin{axcol}
\rnm{f}(\ep) = \ep                                      & \axiom{RN1} \\
\rnm{f}(\dead) = \dead                                  & \axiom{RN2} \\
\rnm{f}(\alpha) = f(\alpha)                             & \axiom{RN3} \\
\rnm{f}(x \altc y) = \rnm{f}(x) \altc \rnm{f}(y)        & \axiom{RN4} \\
\rnm{f}(x \seqc y) = \rnm{f}(x) \seqc \rnm{f}(y)        & \axiom{RN5} \\
\rnm{f}(\phi \gc x) = \phi \gc \rnm{f}(y)               & \axiom{RN6} \\
\rnm{f}(\tau) = \tau                                    & \axiom{RN7}
\end{axcol}
\end{eqntbl}
\end{table}
In this table, 
$f$ stands for an arbitrary function $\funct{f}{\AProcTerm}{\AProcTerm}$ 
such that $f(\alpha) = \alpha$ for all $\alpha \in \AProcASS$,
$\alpha$ stands for an arbitrary term from $\AProcTerm$, and
$\phi$ stands for an arbitrary term from~$\CondTerm$.

Let $t$ be a closed term of the extended theory.
Then the action renaming operator $\rnm{f}$ can be explained as
follows: $\rnm{f}(t)$ denotes the process that behaves the same as the 
process denoted by $t$ except that, where the latter process performs an 
action $\alpha$, the former process performs the action $f(\alpha)$.

Let $T$ be either \deACPet\ or one of its extensions introduced before.
Then we will write $T$+\RN\ for $T$ extended with the action renaming
operators $\rnm{f}$ and the axioms RN1--RN7 from Table~\ref{axioms-rnm}.

\section{Computation and the RAM Conditions}
\label{sect-Computation}

In order to investigate whether \deACPetr\ can play a role in the field 
of models of computation, it has to be explained in the setting of 
\deACPetr\ what it means that a given process computes a given function.
This requires that assumptions about $\gD$ have to be made.
The assumptions concerned are given in this section.
They are based on the idea that the data environment of a computational 
process consists of one or more RAM (Random Access Machine) memories.
Because the assumptions amount to conditions to be satisfied by $\gD$,  
they are called the RAM conditions on $\gD$.
It is also made precise in this section what it means, in the 
setting of \deACPetr\ where $\gD$ satisfies the RAM conditions, that a 
given process computes a given partial function from ${(\BitStr)}^n$ to 
$\BitStr$ ($n \in \Nat$).

\subsection{The RAM Conditions}
\label{subsect-RAM-Conds}

The memory of a RAM consists of a countably infinite number of registers 
which are numbered by natural numbers.
Each register is capable of containing a bit string of arbitrary length.
The contents of the registers constitute the state of the memory of the
RAM.
The execution of an instruction by the RAM amounts to carrying out an 
operation on its memory state that changes the content of at most 
one register or to testing a property of its memory state.
The RAM conditions are presented in this section using the notions of a 
RAM memory state, a RAM operation, and a RAM property.

A \emph{RAM memory state} is a function 
$\funct{\sigma}{\Nat}{\BitStr}$ 
that satisfies the condition that there exists an $i \in \Nat$ such that, 
for all $j \in \Nat$, $\sigma(i + j) = \emptystr$.%
\footnote{We write $\emptystr$ for the empty bit string.} 
We write $\RMState$ for the set of all RAM memory states.

Let $\sigma$ be a RAM memory state.
Then, for all $i \in \Nat$, $\sigma(i)$ is the content of the register 
with number $i$ in memory state $\sigma$.
The condition on $\sigma$ expresses that the part of the memory that is 
actually in use remains finite.

The \emph{input region} and \emph{output region} of a function 
$\funct{o}{\RMState}{\RMState}$, written $\ireg(o)$ and $\oreg(o)$,
respectively, are the subsets of $\Nat$ defined as follows:
\pagebreak[2]
\begin{ldispl}
\oreg(o) =
\Set{i \in \Nat \where
      \Lexists{\sigma \in \RMState}{\sigma(i) \neq o(\sigma)(i)}}\;,
\eqnsep
\ireg(o) =
\Set{i \in \Nat \where
      \Lexists{\sigma_1,\sigma_2 \in \RMState}{{}}
       (\Lforall{j \in \Nat \Sdiff \Set{i}}
         {\sigma_1(j) = \sigma_2(j)} \land {}
\\
\phantom{\ireg(o) = 
         \{i \in \Nat \where \Lexists{\sigma_1,\sigma_2 \in \RMState}{(}}
       \Lexists{j \in \oreg(o)}{o(\sigma_1)(j) \neq o(\sigma_2)(j)})}\;.
\end{ldispl}%

Let $\funct{o}{\RMState}{\RMState}$.
Then $\oreg(o)$ consists of the numbers of all registers that can be 
affected by $o$; and $\ireg(o)$ consists of the numbers of all registers 
that can affect the registers whose numbers are in $\oreg(o)$ under $o$.

A \emph{basic RAM operation} is a function 
$\funct{o}{\RMState}{\RMState}$ that satisfies the condition that 
$\ireg(o)$ is finite and $\oreg(o)$ has cardinality $0$ or $1$.
We write $\RMOp$ for the set of all basic RAM operations.

Let $o$ be a basic RAM operation and $\sigma$ be a RAM memory state.
Then carrying out $o$ on a RAM memory in state $\sigma$ changes the 
state of the RAM memory into $o(\sigma)$.
The condition on $o$ expresses that the content of at most one register 
can be affected and that, if there is such a register, only a finite 
number of registers can affect it. 

The following theorem states that each basic RAM operation transforms 
states of a RAM memory that coincide on its input region to states that 
coincide on its output region.
\begin{theorem} 
\label{theorem-ireg-oreg}
Let $\sigma_1,\sigma_2 \in \RMState$ and
$o \in \RMOp$.
Then $\sigma_1 \restr \ireg(o) = \sigma_2 \restr \ireg(o)$ implies
$o(\sigma_1) \restr \oreg(o) = o(\sigma_2) \restr \oreg(o)$.%
\footnote {We use the notation $f \restr D$, where $f$ is a function and
$D \subseteq \dom(f)$, for the function $g$ with $\dom(g) = D$ such that
 for all $d \in \dom(g)$,\, $g(d) = f(d)$.}
\end{theorem}
\begin{proof}
It is easy to see that the 4-tuple $(\Nat,\BitStr,\RMState,\RMOp)$ is a 
computer according to Definition~3.1 from~\cite{Mau06a}.
From this and Theorem~3.1 from~\cite{Mau06a}, the theorem follows 
immediately.
\qed
\end{proof}

The \emph{input region} of a function $\funct{p}{\RMState}{\Bit}$, 
written $\ireg(p)$ is the subset of $\Nat$ defined as follows:
\begin{ldispl}
\ireg(p) =
\Set{i \in \Nat \where
      \Lexists{\sigma_1,\sigma_2 \in \RMState}{{}}
       (\Lforall{j \in \Nat \Sdiff \Set{i}}
         {\sigma_1(j) = \sigma_2(j)} \land {}
\\
\phantom{\ireg(o) = 
         \{i \in \Nat \where \Lexists{\sigma_1,\sigma_2 \in \RMState}{(}}
        p(\sigma_1) \neq p(\sigma_2)}\;.
\end{ldispl}%

Let $\funct{p}{\RMState}{\Bit}$.
Then $\ireg(p)$ consists of the numbers of all registers that can affect 
what the value of $p$ is.

A \emph{basic RAM property} is a function $\funct{p}{\RMState}{\Bit}$ that 
satisfies the condition that $\ireg(p)$ is finite.
We write $\RMProp$ for the set of all basic RAM properties.

Let $p$ be a basic RAM property and $\sigma$ be a RAM memory state.
Then testing the property $p$ on a RAM memory in state $\sigma$ yields 
the value $p(\sigma)$ and does not change the state of the RAM memory.
The condition on $p$ expresses that only a finite number of registers 
can affect what this value is. 
We say that $p$ \emph{holds in} $\sigma$ if $p(\sigma) = 1$.

The following theorem states that each basic RAM property holds in some 
state of a RAM memory if and only if it holds in all states of the RAM 
memory that coincide with that state on its input region. 
\begin{theorem} 
\label{theorem-ireg-val}
Let $\sigma_1,\sigma_2 \in \RMState$ and
$p \in \RMProp$.
Then $\sigma_1 \restr \ireg(p) = \sigma_2 \restr \ireg(p)$ implies
$p(\sigma_1) = p(\sigma_2)$.
\end{theorem}
\begin{proof}
Let $\RMState'$ be the set of all functions
$\funct{\sigma}{\Nat \Sunion \Set{-1}}{\BitStr}$, 
let $\RMOp'$ be the set of all functions
$\funct{o}{\RMState'}{\RMState'}$,
let $\sigma'_1,\sigma'_2 \in \RMState'$ be such that
$\sigma'_1 \restr \Nat = \sigma_1$ and 
$\sigma'_2 \restr \Nat = \sigma_2$, and
let $o \in \RMOp'$ be such that, 
for all $\sigma \in\RMState'$, 
$o(\sigma) \restr \Nat = \sigma \restr \Nat$ and
$o(\sigma)(-1) = p(\sigma)$.
Then 
$\sigma_1 \restr \ireg(p) = \sigma_2 \restr \ireg(p)$ implies
$p(\sigma_1) = p(\sigma_2)$ iff
$\sigma'_1 \restr \ireg(o) = \sigma'_2 \restr \ireg(o)$ implies
$o(\sigma'_1) = o(\sigma'_2)$.
Because of this and the fact that 
$(\Nat \Sunion \Set{-1},\BitStr,\RMState',\RMOp')$ is also a computer 
according to Definition~3.1 from~\cite{Mau06a}, this theorem now
follows immediately from Theorem~3.1 from~\cite{Mau06a}.
\qed
\end{proof}

With basic RAM operations only computational processes can be considered
whose data environment consists of one RAM memory.
Below, $n$-RAM operations are introduced to remove this restriction.
They are defined such that the basic RAM operations are exactly the 
$1$-RAM operations.

An \emph{$n$-RAM operation} ($n \in \Natpos$) is a function 
$\funct{o}{\RMState^n}{\RMState}$ that satisfies the condition that 
there exist a basic RAM operation $o'$ and 
a $k \in \Nat$ with $1 \leq k \leq n$ such that, 
for all $\sigma_1,\ldots,\sigma_n \in \RMState$,
$o'(\beta(\sigma_1,\ldots,\sigma_n)) =
 \beta(\sigma_1,\ldots,\sigma_{k-1},o(\sigma_1,\ldots,\sigma_n),
       \sigma_{k+1},\ldots,\sigma_n)$, 
where $\funct{\beta}{\RMState^n}{\RMState}$ is the unique function such 
that $\beta(\sigma_1,\ldots,\sigma_n)(n \cdot i + k - 1) = \sigma_k(i)$
for all $i \in \Nat$ and $k \in \Nat$ with $1 \leq k \leq n$.
We write $\nRMOp{n}$, where $n \in \Natpos$, for the set of all $n$-RAM 
operations.

The function 
$\funct{\zeta}{\Set{k \in \Natpos \where k \leq n} \Sx \Nat}{\Nat}$ 
defined by $\zeta(k,i) = n \cdot i + k - 1$ is a bijection.
From this it follows that the basic RAM operation $o'$ and the 
$k \in \Nat$ referred to in the above definition are unique if they 
exist.

The operations from $\Union_{n \geq 1} \nRMOp{n}$ will be referred to as
\emph{RAM operations}.

In a similar way as $n$-RAM operations, $n$-RAM properties are defined.
The basic RAM properties are exactly the $1$-RAM properties.

An \emph{$n$-RAM property} ($n \in \Natpos$) is a function 
$\funct{p}{\RMState^n}{\Bit}$ that satisfies the condition that 
there exists a basic RAM property $p'$ such that, 
for all $\sigma_1,\ldots,\sigma_n \in \RMState$,
$p'(\beta(\sigma_1,\ldots,\sigma_n)) = p(\sigma_1,\ldots,\sigma_n)$, 
where $\funct{\beta}{\RMState^n}{\RMState}$ is defined as above.
We write $\nRMProp{n}$, where $n \in \Natpos$, for the set of all 
$n$-RAM properties.

The properties from $\Union_{n \geq 1} \nRMProp{n}$ will be referred to 
as \emph{RAM properties}.

The RAM conditions on $\gD$ are:
\begin{enumerate}
\item
the signature $\sign_\gD$ of $\gD$ includes:
\begin{itemize}
\item
a sort $\BS$ of \emph{bit strings} and
a sort $\NN$ of \emph{natural numbers};
\item
constants $\const{\emptystr,0,1}{\BS}$ and
a binary operator $\funct{\conc}{\BS \Sx \BS}{\BS}$;
\item
constants $\const{0,1}{\NN}$ and
a binary operator $\funct{+}{\NN \Sx \NN}{\NN}$;
\item
a constant $\const{\ims}{\Data}$ and
a ternary operator $\funct{\upd}{\Data \Sx \NN \Sx \BS}{\Data}$;
\end{itemize}
\item
the sorts, constants, and operators mentioned under 1 are interpreted in
$\gD$ as follows:
\begin{itemize}
\item
the sort $\BS$ is interpreted as the set $\BitStr$,
the sort $\NN$ is interpreted as the set $\Nat$, and
the sort $\Data$ is interpreted as the set $\RMState$;
\item
the constant $\const{\emptystr}{\BS}$ is interpreted as the empty bit 
string,
the constants $\const{0,1}{\BS}$ are interpreted as the bit strings with 
the bit $0$ and $1$, respectively, as sole element, and
the operator $\funct{\conc}{\BS \Sx \BS}{\BS}$ is interpreted as the 
concatenation operation on $\BitStr$;
\item
the constants $\const{0,1}{\NN}$ are interpreted as the natural numbers
$0$ and $1$, respectively, and
the operator $\funct{+}{\NN \Sx \NN}{\NN}$ is interpreted as the addition 
operation on $\Nat$;
\item
the constant $\const{\ims}{\Data}$ is interpreted as the unique 
$\sigma \in \RMState$ such that $\sigma(i) = \emptystr$ for all 
$i \in \Nat$ and
the operator $\funct{\upd}{\Data \Sx \NN \Sx \BS}{\Data}$ is interpreted 
as the \emph{override} operation defined by $\upd(\sigma,i,w)(i) = w$ 
and, for all $j \in \Nat$ with $i \neq j$, 
$\upd(\sigma,i,w)(j) = \sigma(j)$;
\end{itemize}
\item
the signature $\sign_\gD$ of $\gD$ is restricted as follows:
\begin{itemize}
\item
for each operator from $\sign_\gD$, the sort of its 
result is $\Data$ only if the sort of each of its arguments is $\Data$
or the operator is $\upd$;
\item
for each operator from $\sign_\gD$, the sort of its result is $\Bool$ 
only if the sort of each of its arguments is $\Data$;
\end{itemize}
\item
the interpretation of the operators mentioned under 3 is restricted as 
follows:
\begin{itemize}
\item
each operator with result sort $\Data$ other than $\upd$ is interpreted 
as a RAM operation;
\item
each operator with result sort $\Bool$ is interpreted as a RAM property.
\end{itemize}
\end{enumerate}
The notation $\sigma \update{i}{w}$, 
where $\sigma \in \RMState$, $i \in \Nat$, and $w \in \BitStr$,
is used for the term $\upd(\sigma,i,w)$.

The RAM conditions make it possible to explain what it means that a 
given process computes a given partial function from ${(\BitStr)}^n$ to 
$\BitStr$ ($n \in \Nat$).
Moreover, the RAM conditions are nonrestrictive: presumably they allow 
to deal with all proposed versions of the RAM model of computation as 
well as all proposed models of parallel computation that are based on a 
version of the RAM model and the idea that the data environment of a 
computational process consists of one or more RAM memories.

\subsection{Computing Partial Functions from 
\protect$(\BitStr)^n$ to $\BitStr$}
\label{subsect-computability}

In this section, we make precise in the setting of \deACPetrf, where 
$\gD$ is assumed to satisfy the RAM conditions, what it means that a 
given process computes a given partial function from $(\BitStr)^n$ to 
$\BitStr$ ($n \in \Nat$).

In the rest of this paper, $\gD$ is assumed to satisfy the RAM 
conditions.
Moreover, it is assumed that $\RM \in \FlexVar$.

Henceforth, the notation $\rho_{w_1,\ldots,w_n}$, 
where $w_1,\ldots,w_n \in \BitStr$, 
is used for the unique $\rho \in \FVarVal$ such that
$\rho(\RM) = \ims \update{1}{w_1} \ldots \update{n}{w_n}$ and 
$\rho(v) = \ims$ for all $v \in \FlexVar \Sdiff \Set{\RM}$.
%

If $t \in \ProcTermr$ is a finite process term, then there is a finite
upper bound to the number of actions that the process denoted by $t$ can 
perform.

The \emph{depth} of a finite process term $t \in \ProcTermr$, written 
$\depth(t)$, is defined as follows: 
for all $t \in \ProcTerm$, 
$\depth(t) = \min \Set{n \in \Nat \where \proj{n}(t) = t}$.
This means that $\depth(t)$ is the maximum number of actions other than 
$\tau$ that the process denoted by $t$ can perform.

Let $t \in \ProcTermr$,
let $n \in \Nat$, let $\pfunct{F}{{(\BitStr)}^n}{\BitStr}$,%
\footnote
{We write $\pfunct{f}{A}{B}$, where $A$ and $B$ are sets, to indicate 
 that $f$ is a partial function from $A$ to $B$.}
and
let $\funct{W}{\Nat}{\Nat}$. \linebreak[2]
Then $t$ \emph{computes $F$ in $W$ steps} if:
\begin{itemize}
\item
for all $w_1,\ldots,w_n \in \BitStr$ such that $F(w_1,\ldots,w_n)$ 
is defined, there exists a $\sigma \in \RMState$ with 
$\sigma(0) = F(w_1,\ldots,w_n)$ such that:
\begin{ldispl}
\eval{\rho_{w_1,\ldots,w_n}}(t)\; 
\mathrm{is\; a\; terminating\; process\; term}\;,
\seqnsep
\deACPetrf \Ent 
\eval{\rho_{w_1,\ldots,w_n}}(t) =
\eval{\rho_{w_1,\ldots,w_n}}(t \seqc (\RM = \sigma \gc \ep))\;,
\seqnsep 
\depth(\eval{\rho_{w_1,\ldots,w_n}}(t)) \leq 
W(\len(w_1) + \ldots + \len(w_n))\;;%
\footnotemark 
\end{ldispl}%
\footnotetext
{We write $\len(u)$, where $u$ is a sequence, for the length of $u$.}%
\item
for all $w_1,\ldots,w_n \in \BitStr$ such that $F(w_1,\ldots,w_n)$ is
undefined, 
for all $\rho \in \FVarVal$ with 
$\rho(\RM) = \ims \update{1}{w_1} \ldots \update{n}{w_n}$:
\begin{ldispl}
\eval{\rho_{w_1,\ldots,w_n}}(t)\; 
\mathrm{is\; not\; a\; terminating\; process\; term}\;. 
\end{ldispl}%
\end{itemize}
We say that $t$ \emph{computes $F$} if there exists a 
$\funct{W}{\Nat}{\Nat}$ such that $t$ computes $F$ in $W$ steps,
we say that $F$ \emph{is a computable function} if there exists a
$t \in \ProcTerm$ such that $t$ computes $F$, and
we say that $t$ \emph{is a computational process} if there exists a 
$\pfunct{F}{{(\BitStr)}^n}{\BitStr}$ such that $t$ computes $F$.

We write $\CProcTermr$ for the set
$\Set{t \in \ProcTermr \where
 t \mathrm{\;is\; a\; computational\; process}}$.

With the above definition, we can establish whether a process of the 
kind considered in the current setting computes a given partial function 
from ${(\BitStr)}^n$ to $\BitStr$ ($n \in \Nat$) by equational reasoning 
using the axioms of \linebreak[2] \deACPetrf.
This setting is more general than the setting provided by any known 
version of the RAM model of computation.
It is not suitable as a model of computation itself.
However, various known models of computation can be defined by fixing 
which RAM operations and which RAM properties belong to $\gD$ and by 
restricting the computational processes to the ones of a certain form. 
To the best of my knowledge, the models of computation that can be dealt 
with in this way include all proposed versions of the RAM model as well 
as all proposed models of parallel computation that are based on a 
version of the RAM model and the idea that the data environment of a 
computational process consists of one or more RAM memories.

Whatever model of computation is obtained by fixing the RAM operations 
and the RAM properties and by restricting the computational processes to 
the ones of a certain form, it is an idealization of a real computer 
because it offers an unbounded number of registers that can contain a 
bit string of arbitrary length instead of a bounded number of registers 
that can only contain a bit string of a fixed length.

\section{The \RAMP\ Model of Computation}
\label{sect-RAM-Model}

In this section, a version of the RAM model of computation is described
in the setting introduced in the previous sections.
Because it focuses on the processes that are produces by RAMs when they
execute their built-in program, the version of the RAM model of 
computation described in this section is called the \RAMP\ (Random 
Access Machine Process) model of computation. 

First, the operators are introduced that represent the RAM operations 
and the RAM properties that belong to $\gD$ in the case of the \RAMP\ 
model of computation.
Next, the interpretation of those operators as a RAM operation or a RAM 
property is given. 
Finally, the \RAMP\ model of computation is described.

\subsection{Operators for the \RAMP\ Model}
\label{subsect-RAM-Model-Operators}

In this section, the operators that are relevant to the \RAMP\ model of
computation are introduced.

In the case of the \RAMP\ model of computation, the set of operators 
from $\sign_\gD$ that are interpreted in $\gD$ as RAM operations or RAM 
properties is the set $\RAMOp$ defined as follows:
\begin{ldispl}
\begin{aeqns}
\RAMOp & = &
\Set{\binop{:}\src_1{:}\src_2{:}\dst \where
     \binop \in \Binop \Land \src_1,\src_2 \in Src \Land \dst \in \Dst}
\\ & \Sunion &
\Set{\unop{:}\src_1{:}\dst \where
     \unop \in \Unop \Land  \src_1 \in Src \Land \dst \in \Dst}
\\ & \Sunion &
\Set{\cmpop{:}\src_1{:}\src_2 \where
     \cmpop \in \Cmpop \Land \src_1,\src_2 \in Src}\;,
\end{aeqns}
\end{ldispl}%
where 
\begin{ldispl}
\begin{aeqns}
\Binop & = & \Set{\addop,\subop,\andop,\orop}\;,
\\
\Unop  & = & \Set{\notop,\shlop,\shrop,\movop}\;,
\\
\Cmpop & = & \Set{\eqop,\gtop,\beqop}\;
\end{aeqns}
\end{ldispl}%
and
\begin{ldispl}
\begin{aeqns}
\Src & = &
\Set{\# i \where i \in \Nat} \Sunion \Nat \Sunion
\Set{@ i \where i \in \Nat}\;,
\\
\Dst & = & \Nat \Sunion \Set{@ i \where i \in \Nat}\;.
\end{aeqns}
\end{ldispl}%
We write $\RAMOpp$ for the set
$\Set{\cmpop{:}\src_1{:}\src_2 \where
      \cmpop \in \Cmpop \Land \src_1,\src_2 \in Src}$ and 
$\RAMOpo$ for the set $\RAMOp \Sdiff \RAMOpp$.

The operators from $\RAMOpo$ are the operators that are 
interpreted in $\gD$ as basic RAM operations and the operators from 
$\RAMOpp$  are the operators that are interpreted in $\gD$ as basic RAM 
properties.

The following is a preliminary explanation of the operators from $\RAMOp$:
\begin{itemize}
\item
carrying out the operation denoted by an operator of the form 
$\binop{:}\src_1{:}\src_2{:}\dst$ on a RAM memory in some state boils 
down to carrying out the binary operation named $\binop$ on the values 
that $\src_1$ and $\src_2$ stand for in that state and then changing 
the content of the register that $\dst$ stands for into the result of 
this;
\item
carrying out the operation denoted by an operator of the form 
$\unop{:}\src_1{:}\dst$ \linebreak[2] on a RAM memory in some state boils 
down to carrying out the unary operation named $\unop$ on the value that 
$\src$ stands for in that state and then changing the content of the 
register that $\dst$ stands for into the result of this;
\item
carrying out the operation denoted by an operator of the form 
$\cmpop{:}\src_1{:}\src_2$ on a RAM memory in some state boils down to 
carrying out the binary operation named $\cmpop$ on the values 
that $\src_1$ and $\src_2$ stand for in that state.
\end{itemize}
The value that $\src_i$ ($i = 1,2$) stands for is as follows:
\begin{itemize}
\item
\emph{immediate}: it stands for the shortest bit string representing the 
natural number $i$ if it is of the form $\# i$;
\item
\emph{direct addressing}: it stands for the content of the register with 
number $i$ if it is of the form $i$;
\item
\emph{indirect addressing}: it stands for the content of the register 
whose number is represented by the content of the register with number 
$i$ if it is of the form~$@ i$;
\end{itemize}
and the register that $\dst$ stands for is as follows:
\begin{itemize}
\item
\emph{direct addressing}: it stands for the register with number $i$ if 
it is of the form~$i$;
\item
\emph{indirect addressing}: it stands for the register whose number is 
represented by the content of the register with number $i$ if it is of 
the form $@ i$.
\end{itemize}

The following kinds of operations and relations on bit strings are 
covered by the operators from $\RAMOp$: 
\emph{arithmetic} operations ($\addop,\subop$),
\emph{logical} operations ($\andop,\orop,\notop$),
\emph{bit-shift} operations ($\shlop,\shrop$), 
\emph{data-transfer} operations ($\movop$), 
\emph{arithmetic} relations ($\eqop,\gtop$), and
the bit-wise equality relation ($\beqop$).
The arithmetic operations on bit strings are operations that model 
arithmetic operations on natural numbers with respect to their binary 
representation by bit strings,
the logical operations on bit strings are bitwise logical operations,
and the data transfer operation on bit strings is the identity operation 
on bit strings (which is carried out when copying bit strings).
The arithmetic relations on bit strings are relations that model 
arithmetic relations on natural numbers with respect to their binary 
representation by bit strings.

\subsection{Interpretation of the Operators for the \RAMP\ Model}
\label{subsect-Interpretation-Operators}

In this section, the interpretation of the operators from $\RAMOp$ in 
$\gD$ is defined.

We start with defining auxiliary functions for conversion between 
natural numbers and bit strings and evaluation of the elements of $\Src$ 
and $\Dst$. 

We write $\monus$ for proper subtraction of natural numbers.
We write $\div$ for zero-totalized Euclidean division of natural 
numbers, i.e.\ Euclidean division made total by imposing that division 
by zero yields zero (like in meadows, see \mbox{e.g.~\cite{BT07a,BM09g}}).
We use juxtaposition for concatenation of bit strings.

The \emph{natural to bit string} function 
$\funct{\ntob}{\Nat}{\BitStr}$ 
is recursively defined as follows:
\begin{itemize}
\item[] 
$\ntob(n) = n$ if $n \leq 1 \quad$ and 
$\quad \ntob(n) = (n \bmod 2) \cat \ntob(n \div 2)$ if $n > 1$
\end{itemize}
and the \emph{bit string to natural} function 
$\funct{\bton}{\BitStr}{\Nat}$ 
is recursively defined as follows: 
\begin{itemize}
\item[] 
$\bton(\emptystr) = 0 \quad$ and 
$\quad \bton(b \cat w) = b + 2 \mul \bton(w)$.
\end{itemize}
These definitions tell us that, when viewed as the binary representation 
of a natural number, the first bit of a bit string is considered the 
least significant bit.
Results of applying $\ntob$ have no leading zeros, but the operand of 
$\bton$ may have leading zeros.
Thus, we have that $\bton(\ntob(n)) = n$ and $\ntob(\bton(w)) = w'$, 
where $w'$ is $w$ without leading zeros.

For each $\sigma \in \RMState$, the \emph{src-valuation in $\sigma$} 
function $\funct{\val{\sigma}}{\Src}{\BitStr}$ is defined as follows:
\begin{itemize}
\item[] 
$\val{\sigma}(\# i) = \ntob(i)$, $\val{\sigma}(i) = \sigma(i)$, and
$\val{\sigma}(@ i) = \sigma(\bton(\sigma(i)))$  
\end{itemize}
and, 
for each $\sigma \in \RMState$, the \emph{dst-valuation in $\sigma$} 
function $\funct{\reg{\sigma}}{\Dst}{\Nat}$ is defined as follows:
\begin{itemize}
\item[] 
$\reg{\sigma}(i) = i$ and $\reg{\sigma}(@ i) = \bton(\sigma(i))$. 
\end{itemize}

We continue with defining the operations on bit strings that the 
operation names from $\Binop \Sunion \Unop$ refer to.

We define the operations on bit strings that the operation names
$\addop$ and $\subop$ refer to as follows:
\begin{ldispl}
\begin{tabular}[t]{@{}l@{\,\,}l@{}}
$\funct{+}{\BitStr \Sx \BitStr}{\BitStr}$: &
$w_1 + w_2 = \ntob(\bton(w_1) + \bton(w_2))$;
\\
$\funct{\monus}{\BitStr \Sx \BitStr}{\BitStr}$: &
$w_1 \monus w_2 = \ntob(\bton(w_1) \monus \bton(w_2))$.
\end{tabular}
\end{ldispl}%
These definitions tell us that, although the operands of the operations 
$+$ and~$\monus$ may have leading zeros, results of applying these 
operations have no leading zeros.

We define the operations on bit strings that the operation names
$\andop$, $\orop$, and $\notop$ refer to recursively as follows:
\begin{ldispl}
\begin{tabular}[t]{@{}l@{}}
$\funct{\Land}{\BitStr \Sx \BitStr}{\BitStr}$:\hsp{.25}
$\emptystr \Land \emptystr = \emptystr$,\hsp{.25}
$\emptystr \Land (b \cat w) = \zero \cat (\emptystr \Land w)$, 
\\ \hsp{.75}
$(b \cat w) \Land \emptystr = \zero \cat (w \Land \emptystr)$,\hsp{.25}
$(b_1 \cat w_1) \Land (b_2 \cat w_2) =
 (b_1 \Land b_2) \cat (w_1 \Land w_2)$;
\seqnsep
$\funct{\Lor}{\BitStr \Sx \BitStr}{\BitStr}$:\hsp{.25}
$\emptystr \Lor \emptystr = \emptystr$,\hsp{.25}
$\emptystr \Lor (b \cat w) = b \cat (\emptystr \Lor w)$, \\ \hsp{.75}
$(b \cat w) \Lor \emptystr = b \cat (w \Lor \emptystr)$,\hsp{.25}
$(b_1 \cat w_1) \Lor (b_2 \cat w_2) =
 (b_1 \Lor b_2) \cat (w_1 \Lor w_2)$;
\seqnsep
$\funct{\Lnot}{\BitStr}{\BitStr}$:\hsp{.25}
$\Lnot \emptystr = \emptystr$,\hsp{.25}
$\Lnot (b \cat w) = (\Lnot b) \cat (\Lnot w)$.
\end{tabular}
\end{ldispl}%
These definitions tell us that, if the operands of the operations 
$\Land$ and $\Lor$ do not have the same length, sufficient leading zeros 
are assumed to exist.
Moreover, results of applying these operations and results of applying 
$\Lnot$ can have leading zeros.

We define the operations on bit strings that the operation names 
$\shlop$ and $\shrop$ refer to as follows:
\begin{ldispl}
\begin{tabular}[t]{@{}l@{\,\,}l@{}}
$\funct{\shl}{\BitStr}{\BitStr}$: & 
$\shl \emptystr = \emptystr$,\hsp{.25} 
$\shl (b \cat w) = \zero \cat b \cat w$;
\\
$\funct{\shr}{\BitStr}{\BitStr}$: &
$\shr \emptystr = \emptystr$,\hsp{.25} 
$\shr (b \cat w) = w$.
\end{tabular}
\end{ldispl}%
These definitions tell us that results of applying the operations $\shl$ 
and $\shr$ can have leading zeros.
We have that $\bton(\shl w) = \bton(w) \mul 2$ and 
$\bton(\shr w) = \bton(w) \div 2$.

Now, we are ready to define the interpretation of the operators from 
$\RAMOp$ in $\gD$.
For each $o \in \RAMOp$, the interpretation of $o$ in $\gD$, written
$\Int{o}$, is defined as follows:
\begin{ldispl}
\begin{tabular}[t]{@{}l@{\,\,}c@{\,\,}l@{}}
$\Int{\addop{:}s_1{:}s_2{:}d}(\sigma)$ & = & $
 \sigma\update{\reg{\sigma}(d)}{\val{\sigma}(s_1) + \val{\sigma}(s_2)}$;
\\
$\Int{\subop{:}s_1{:}s_2{:}d}(\sigma)$ & = & $
 \sigma\update{\reg{\sigma}(d)}
  {\val{\sigma}(s_1) \monus \val{\sigma}(s_2)}$;
\\
$\Int{\andop{:}s_1{:}s_2{:}d}(\sigma)$ & = & $
 \sigma\update{\reg{\sigma}(d)}
  {\val{\sigma}(s_1) \Land \val{\sigma}(s_2)}$;
\\
$\Int{\orop{:}s_1{:}s_2{:}d}(\sigma)$ & = & $
 \sigma\update{\reg{\sigma}(d)}
  {\val{\sigma}(s_1) \Lor \val{\sigma}(s_2)}$;
\\
$\Int{\notop{:}s_1{:}d}(\sigma)$ & = & $
 \sigma\update{\reg{\sigma}(d)}{\Lnot \val{\sigma}(s_1)}$;
\\
$\Int{\shlop{:}s_1{:}d}(\sigma)$ & = & $
 \sigma\update{\reg{\sigma}(d)}{\shl \val{\sigma}(s_1)}$;
\\
$\Int{\shrop{:}s_1{:}d}(\sigma)$ & = & $
 \sigma\update{\reg{\sigma}(d)}{\shr \val{\sigma}(s_1)}$;
\\
$\Int{\movop{:}s_1{:}d}(\sigma)$ & = & $
 \sigma\update{\reg{\sigma}(d)}{\val{\sigma}(s_1)}$;
\\[.5ex] 
$\Int{\eqop{:}s_1{:}s_2}(\sigma)$ & = & $ 
\left \{
\begin{array}{l}
1 \;\mathrm{if}\; \bton(\val{\sigma}(s_1)) = \bton(\val{\sigma}(s_2)),
\\
0 \;\mathrm{otherwise};
\end{array}
\right.
$
\\[2ex]
$\Int{\gtop{:}s_1{:}s_2}(\sigma)$ & = & $ 
\left \{
\begin{array}{l}
1 \;\mathrm{if}\; \bton(\val{\sigma}(s_1)) > \bton(\val{\sigma}(s_2)),
\\
0 \;\mathrm{otherwise};
\end{array}
\right.
$
\\[2ex]
$\Int{\beqop{:}s_1{:}s_2}(\sigma)$ & = & $
\left \{
\begin{array}{l}
1 \;\mathrm{if}\; \val{\sigma}(s_1) = \val{\sigma}(s_2),
\\
0 \;\mathrm{otherwise}.
\end{array}
\right.
$
\end{tabular}
\end{ldispl}

Clearly, the interpretation of each operator from $\RAMOpo$
is a basic RAM operation and the interpretation of each operator from 
$\RAMOpp$ is a basic RAM property.

\subsection{\RAMP\ Terms and the \RAMP\ Model of Computation}
\label{subsect-RAM-processes}

In this section, the \RAMP\ model of computation is characterized in the 
setting introduced in Sections~\ref{sect-deACPet} 
and~\ref{sect-Computation}.
However, first the notion of a \RAMP\ term is defined. 
This notion is introduced to make precise what the set of possible 
computational processes is in the case of the \RAMP\ model of 
computation.

In this section, $\gD$ is fixed as follows:
\begin{itemize}
\item
$\sign_\gD$ is the smallest signature including
(a)~all sorts, constants, and operators required by the assumptions made 
about $\gD$ in \deACPet\ or the RAM conditions on $\gD$ and
(b)~all operators from $\RAMOp$;
\item
all sorts, constants, and operators mentioned under~(a) are interpreted 
in $\gD$ as required by the assumptions made about $\gD$ in \deACPet\ or 
the RAM conditions on $\gD$;
\item
all operators mentioned under~(b) are interpreted in $\gD$ as defined at 
the end of Section~\ref{subsect-Interpretation-Operators}.
\end{itemize}
Moreover, it is assumed that $\RM \in \FlexVar$.

A \emph{RAM process term}, called a \emph{\RAMP\ term} for short, is a 
term from $\ProcTermr$ that is of the form $\rec{X}{E}$, where, for each 
$Y \in \vars(E)$, the recursion equation for $Y$ in $E$ has one of the 
following forms:
\begin{ldispl}
\begin{tabular}[t]{@{}l@{}}
$Y = \True \gc \ass{\RM}{o(\RM)} \seqc Z$, 
\\
$Y = (p(\RM) = 1) \gc \ass{\RM}{\RM} \seqc Z \altc 
     (p(\RM) = 0) \gc \ass{\RM}{\RM} \seqc Z'$, 
\\
$Y = \True \gc \ep$,
\end{tabular}
\end{ldispl}%
where $o \in \RAMOpo$, $p \in \RAMOpp$, and 
$Z,Z' \in \vars(E)$.
We write $\RAMProcTerm$ for the set of all \RAMP\ terms, and
we write $\CRAMProcTerm$ for $\RAMProcTerm \Sinter \CProcTermr$.

A process that can be denoted by a \RAMP\ term is called a
\emph{RAM process} or a \emph{\RAMP}\ for short.
So, a \RAMP\ is a process that is definable by a guarded linear 
recursive specification over \deACPet\ of the kind described above.

As mentioned in Section~\ref{sect-intro}, a basic assumption in this 
paper is that a model of computation is fully characterized by:
(a)~a set of possible computational processes,
(b)~for each possible computational process, a set of possible data 
environments, and
(c)~the effect of applying such processes to such environments.

$\gD$ as fixed above and $\CRAMProcTerm$ induce the \RAMP\ model of 
computation: 
\begin{itemize}
\item
the set of possible computational processes is the set of all processes
that can be denoted by a term from $\CRAMProcTerm$;
\item
for each possible computational process, the set of possible data 
environments is the set of all $\Set{\RM}$-indexed data environments; 
\item
the effect of applying the process denoted by a $t \in \CRAMProcTerm$ to 
a $\Set{\RM}$-indexed data environment $\mu$ is $\eval{\rho}(t)$, where 
$\rho$ is a flexible variable valuation that represents $\mu$.
\end{itemize}

The \RAMP\ model of computation described above is intended to be 
essentially the same as the standard RAM model of computation extended 
with logical instructions and bit-shift instructions.
The RAMs from that model will be referred to as the BBRAMs (Basic Binary 
RAMs).
There is a strong resemblance between $\RAMOp$ and the set $\RAMInstr$
of instructions from which the built-in programs of the BBRAMs can be 
constructed.
Because the concrete syntax of the instructions does not matter, 
$\RAMInstr$ can be defined as follows:
\begin{ldispl}
\RAMInstr = 
(\RAMOpo) \Sunion
\Set{\jmpinstr{:}p{:}i \where p \in \RAMOpp \Land i \in \Natpos} \Sunion
\Set{\haltinstr}\;.
\end{ldispl}%
A \emph{BBRAM program} is a non-empty sequence $C$ from $\RAMInstr^*$ in 
which instructions of the form $\jmpinstr{:}p{:}i$ with $i > \len(C)$ do 
not occur.
We write $\RAMProg$ for the set of all BBRAM programs.

The execution of an instruction $o$ from $\RAMOpo$ by a 
BBRAM causes the state of its memory to change according to $\Int{o}$. 
The execution of an instruction of the form $\jmpinstr{:}p{:}i$ or the
instruction $\haltinstr$ by a BBRAM has no effect on the state of its
memory.
After execution of an instruction by a BBRAM, the BBRAM proceeds to the 
execution of the next instruction from its built-in program except when 
the instruction is of the form $\jmpinstr{:}p{:}i$ and $\Int{p} = 1$ or
when the instruction is $\haltinstr$.
In the case that the instruction is of the form $\jmpinstr{:}p{:}i$ and 
$\Int{p} = 1$, the execution proceeds to the $i$th instruction of the 
program. 
In the case that the instruction is $\haltinstr$, the execution 
terminates successfully.

The processes that are produced by the BBRAMs when they execute their 
built-in program are given by a function 
$\funct{\process}{\RAMProg}{\RAMProcTerm}$ that is defined up to
consistent renaming of variables as follows:
$\process(c_1\, \ldots\, c_n) = \rec{X_1}{E}$, 
where \linebreak[2] $E$ consists of, 
for each $i \in \Nat$ with $1 \leq i \leq n$, an equation
\begin{ldispl}
\begin{tabular}[t]{@{}l@{\,\,}l@{}}
$X_i = \True \gc \ass{\RM}{c_i(\RM)} \seqc X_{i+1}$ &
\hsp{9} if $c_i \in \RAMOpo$, 
\\
\multicolumn{2}{@{}l@{}}
{$X_i = (p(\RM) = 1) \gc \ass{\RM}{\RM} \seqc X_j\altc 
       (p(\RM) = 0) \gc \ass{\RM}{\RM} \seqc X_{i+1}$} \\ &
\hsp{9} if $c_i \equiv \jmpinstr{:}p{:}j$, 
\\
$X_i = \True \gc \ep$ &
\hsp{9} if $c_i \equiv \haltinstr$,
\end{tabular}
\end{ldispl}%
where $X_1,\ldots,X_n$ are different variable from $\cX$.

Let $C \in \RAMProg$.
Then $\process(C)$ denotes the process that is produced by the BBRAM 
whose built-in program is $C$ when it executes its built-in program.

The definition of $\process$ is in accordance with the descriptions of 
various versions of the RAM model of computation in the literature on 
this subject (see e.g.~\cite{CR73a,HS74a,AHU74a,Pap94a}).
However, to the best of my knowledge, none of these descriptions is 
precise and complete enough to allow of a proof of this.

The RAMPs are exactly the processes that can be produced by 
the BBRAMs when they execute their built-in program.
\begin{theorem}
\label{theorem-RAMP-RAM}
For each constant $\rec{X}{E} \in \ProcTermr$, 
$\rec{X}{E} \in \RAMProcTerm$ iff 
there exists \linebreak[2] a $C \in \RAMProg$ such that $\rec{X}{E}$ and 
$\process(C)$ are identical up to consistent renaming of variables.
\end{theorem}
\begin{proof}
It is easy to see that
(a)~for all $C \in \RAMProg$, $\process(C) \in \RAMProcTerm$ and 
(b)~$\process$~is an bijection up to consistent renaming of variables.
From this, the theorem follows immediately.
\qed
\end{proof}
Notice that, if $\rec{X}{E}$ and $\rec{X'}{E'}$ are identical up to 
consistent renaming of variables, then the equation 
$\rec{X}{E} = \rec{X'}{E'}$ is derivable from RDP and RSP (and 
$\rec{X}{E} \rbbisim \rec{X'}{E'}$).

The following theorem is a result concerning the computational power of
RAMPs.
\begin{theorem}
\label{theorem-Turing-computable-RAMP}
For each $\pfunct{F}{{(\BitStr)}^n}{\BitStr}$, there exists a 
$t \in \RAMProcTerm$ such that $t$ computes $F$ iff $F$ is 
Turing-computable.
\end{theorem}
\begin{proof}
By Theorem~\ref{theorem-RAMP-RAM}, it is sufficient to show that each 
BBRAM is Turing equivalent to a Turing machine.
The BBRAM model of computation is essentially the same as the BRAM model 
of computation from~\cite{Emd90a} extended with bit-shift instructions.
It follows directly from simulation results mentioned in~\cite{Emd90a} 
(part~(3) of Theorem~2.4, part~(3) of Theorem~2.5, and part~(2) of 
Theorem~2.6) that each BRAM can be simulated by a Turing machine and 
vice versa.
Because each Turing machine can be simulated by a BRAM, we immediate 
have that each Turing machine can be simulated by a BBRAM.
It is easy to see that the bit-shift instructions can be simulated by 
a Turing machine.
From this and the fact that each BRAM can be simulated by a Turing 
machine, it follows that each BBRAM can be simulated by a Turing 
machine as well. 
Hence, each BBRAM is Turing equivalent to a Turing machine.
\qed
\end{proof}

Henceforth, we write $\POLY$ for 
$\Set{f \where 
 \funct{f}{\Nat}{\Nat} \Land f \mathrm{\,is\,a\,polynomial\,function}}$.
The following theorem tells us that the decision problems that belong to 
$\mathbf{P}$ are exactly the decision problems that can be solved by 
means of a RAMP in polynomially many steps.
\begin{theorem}
\label{theorem-P}
For each $\funct{F}{\BitStr}{\Bit}$, there exist a $t \in \RAMProcTerm$ 
and a \linebreak[2] $W \in \POLY$ such that $t$ computes $F$ in $W$ 
steps iff $F \in \mathbf{P}$.
\end{theorem}
\begin{proof}
By Theorem~\ref{theorem-RAMP-RAM}, it is sufficient to show that time 
complexity on BBRAMs under the uniform time measure, i.e.\ the number of 
steps, and time complexity on multi-tape Turing machines are 
polynomially related.
The BBRAM is essentially the same as the BRAM model of computation 
from~\cite{Emd90a} extended with bit-shift instructions.
It follows directly from simulation results mentioned in~\cite{Emd90a} 
(part~(3) of Theorem~2.4, part~(3) of Theorem~2.5, and part~(2) of 
Theorem~2.6) that time complexity on BRAMs under the uniform time 
measure and time complexity on multi-tape Turing machines are 
polynomially related.
It is easy to see that the bit-shift instructions can be simulated by a 
multi-tape Turing machine in linear time.
Hence, the time complexities remain polynomially related if the BRAM 
model is extended with the bit-shift instructions.
\qed
\end{proof}

\section{The \APRAMP\ Model of Computation}
\label{sect-APRAM-Model}

In this section, an asynchronous parallel RAM model of computation is 
described in the setting introduced in Sections~\ref{sect-deACPet} 
and~\ref{sect-Computation}.
Because it focuses on the processes that are produces by asynchronous 
parallel RAMs when they execute their built-in programs, the parallel 
RAM model of computation described in this section is called the 
\APRAMP\ (Asynchronous Parallel Random Access Machine Process) model of 
computation. 
In this model of computation, a computational process is the parallel 
composition of a number of processes that each has its own private RAM 
memory.
However, together they also have a shared RAM memory for synchronization
and communication.

First, the operators are introduced that represent the RAM operations 
and the RAM properties that belong to $\gD$ in the case of the \APRAMP\ 
model of computation.
Next, the interpretation of those operators as a RAM operation or a RAM 
property is given.
Finally, the \APRAMP\ model of computation is described.

In the case of the \APRAMP\ model of computation, the set of operators 
from $\sign_\gD$ that are interpreted in $\gD$ as RAM operations or RAM 
properties is the set $\APRAMOp$ defined as follows:
\begin{ldispl}
\begin{aeqns}
\APRAMOp & = & \RAMOp \Sunion \Set{\iniop{:}\# i \where i \in \Natpos} \\
 & \Sunion & 
\Set{\loaop{:}@ i{:}\dst \where i \in \Nat \Land \dst \in \Dst} \Sunion 
\Set{\stoop{:}\src{:}@ i \where \src \in Src \Land i \in \Nat}\;,
\end{aeqns}
\end{ldispl}%
where $\Src$ and $\Dst$ are as defined in 
Section~\ref{subsect-RAM-Model-Operators}.

\pagebreak[2]
In operators of the forms $\binop{:}\src_1{:}\src_2{:}\dst$, 
$\unop{:}\src_1{:}\dst$, and $\cmpop{:}\src_1{:}\src_2$ from $\RAMOp$,
$\src_1$, $\src_2$, and $\dst$ refer to the private RAM memory.
In operators of the form $\loaop{:}@ i{:}\dst$ and 
$\stoop{:}\src{:}@ i$ from $\APRAMOp \Sdiff \RAMOp$, $\src$ and $\dst$ 
refer to the private RAM memory too.
The operators of the form $\loaop{:}@ i{:}\dst$ and 
$\stoop{:}\src{:}@ i$ differ from the operators of the form 
$\movop{:}@ i{:}\dst$ and $\movop{:}\src{:}@ i$, respectively, in that 
$@ i$ stands for the content of the register and the register, 
respectively, from the shared RAM memory whose number is represented by 
the content of the register with number $i$ from the private memory.
The operators of the form $\iniop{:}\# i$ initialize the registers from 
the private memory as follows:
the content of the register with number $0$ becomes the shortest bit 
string that represents the natural number $i$ and the content of all 
other registers becomes the empty bit string.

Now, we are ready to define the interpretation of the operators from 
$\APRAMOp$ in $\gD$.
For each $o \in \APRAMOp$, the interpretation of $o$ in $\gD$, written
$\Int{o}$, is as defined in Section~\ref{subsect-Interpretation-Operators}
for operators from $\RAMOp$ and as defined below for the additional 
operators:
\begin{ldispl}
\begin{tabular}[t]{@{}l@{\,\,}c@{\,\,}l@{}}
$\Int{\iniop{:}\# i}(\sigma_p)$ & = & $ \ims\update{0}{\ntob(i)}$;
\\
$\Int{\loaop{:}@ i{:}d}(\sigma_p,\sigma_s)$ & = & $
 \sigma_p\update{\reg{\sigma_p}(d)}{\sigma_s(\bton(\sigma_p(i)))}$;
\\
$\Int{\stoop{:}s{:}@ i}(\sigma_p,\sigma_s)$ & = & $
 \sigma_s\update{\bton(\sigma_p(i))}{\val{\sigma_p}(s)}$.
\end{tabular}
\end{ldispl}

Clearly, the interpretation of each operator of the form $\iniop{:}\# i$ 
is a $1$-RAM operation and the interpretation of each operator of the 
form $\loaop{:}@ i{:}d$ or $\stoop{:}s{:}@ i$ is a $2$-RAM operation.

Below, the \APRAMP\ model of computation is characterized in the setting 
introduced in Sections~\ref{sect-deACPet} and~\ref{sect-Computation}.
However, first the notion of an \APRAMP\ term is defined. 
This notion is introduced to make precise what the set of possible 
computational processes is in the case of the \APRAMP\ model of 
computation.

In this section, $\gD$ is fixed as follows:
\begin{itemize}
\item
$\sign_\gD$ is the smallest signature including
(a)~all sorts, constants, and operators required by the assumptions made 
about $\gD$ in \deACPet\ or the RAM conditions on $\gD$ and
(b)~all operators from $\APRAMOp$;
\item
all sorts, constants, and operators mentioned under~(a) are interpreted 
in $\gD$ as required by the assumptions made about $\gD$ in \deACPet\ or 
the RAM conditions on $\gD$;
\item
all operators mentioned under~(b) are interpreted in $\gD$ as defined 
above.
\end{itemize}

Moreover, it is assumed that $\RM \in \FlexVar$ and, for all 
$i \in \Natpos$, $\RM_i \in \FlexVar$.
We write $\FlexVar^\RM_n$, where $n \in \Natpos$, for the set
$\Set{\RM} \Sunion \Set{\RM_i \where i \in \Natpos \Land i \leq n}$.

An \emph{$n$-APRAM process term} ($n \in \Natpos$), called an 
\emph{$n$-\APRAMP\ term} for short, is a term from $\ProcTermr$ that is 
of the form $\rec{X_1}{E_1} \parc \ldots \parc \rec{X_n}{E_n}$, where, 
for each $i \in \Natpos$ with $i \leq n$: 
\begin{itemize}
\item
for each $X \in \vars(E_i)$, the recursion equation for $X$ in $E_i$ has 
one of the following forms:
\pagebreak[2]
\begin{ldispl}
\begin{tabular}[t]{@{}l@{\;\;}l@{}}
(1) & $X = \True \gc \ass{\RM_i}{\iniop{:}\# i\,(\RM_i)} \seqc Y$, 
\\
(2) & $X = \True \gc \ass{\RM_i}{\loaop{:}@ i{:}d\,(\RM_i,\RM)} \seqc Y$, 
\\
(3) & $X = \True \gc \ass{\RM}{\stoop{:}s{:}@ i\,(\RM_i,\RM)} \seqc Y$, 
\\
(4) & $X = \True \gc \ass{\RM_i}{o(\RM_i)} \seqc Y$, 
\\
(5) & $X = (p(\RM_i) = 1) \gc \ass{\RM_i}{\RM_i} \seqc Y \altc 
       (p(\RM_i) = 0) \gc \ass{\RM_i}{\RM_i} \seqc Y'$, 
\\
(6) & $X = \True \gc \ep$,
\end{tabular}
\end{ldispl}%
where $o \in \RAMOpo$, $p \in \RAMOpp$, and 
$Y,Y' \in \vars(E_i)$;
\item
for each $X \in \vars(E_i)$, the recursion equation for $X$ in $E_i$ is 
of the form~(1) iff $X \equiv X_i$.
\end{itemize}
We write $\APRAMProcTerm$ for the set of all terms $t \in \ProcTermr$
such that $t$ is an $n$-\APRAMP\ terms for some $n \in \Natpos$, and
we write $\CAPRAMProcTerm$ for $\APRAMProcTerm \Sinter \CProcTermr$.
Moreover, we write $\nm{deg}(t)$, where $t \in \APRAMProcTerm$, for the 
unique $n \in \Natpos$ such that $t$ is an $n$-\APRAMP\ term.

The terms from $\APRAMProcTerm$ will be referred to as 
\emph{\APRAMP\ terms}.

A process that can be denoted by an \APRAMP\ term is called an
\emph{APRAM process} or an \emph{\APRAMP}\ for short.
So, an \APRAMP\ is a parallel composition of processes that are 
definable by a guarded linear recursive specification over \deACPet\ of 
the kind described above.
Each of those parallel processes starts with an initialization step in 
which the number of its private memory is made available in the register 
with number $0$ from its private memory.

Notice that by Lemma~\ref{lemma-glr-abstr-free} and
Theorem~\ref{theorem-completeness-ACPet-1},
for all $t \in \APRAMProcTerm$, there exists a guarded linear recursive 
specification $E$ and $X \in \vars(E)$ such that 
$t \rbbisim \rec{X}{E}$.

As mentioned before, a basic assumption in this paper is that a model of 
computation is fully characterized by:
(a)~a set of possible computational processes,
(b)~for each possible computational process, a set of possible data 
environments, and
(c)~the effect of applying such processes to such environments.

$\gD$ as fixed above and $\CAPRAMProcTerm$ induce the \APRAMP\ model of 
computation: 
\begin{itemize}
\item
the set of possible computational processes is the set of all processes
that can be denoted by a term from $\CAPRAMProcTerm$;
\item
for each possible computational process $p$, the set of possible data 
environments is the set of all $\FlexVar^\RM_{\nm{deg}(t)}$-indexed data 
environments, where $t$ is a term from $\CAPRAMProcTerm$ denoting $p$; 
\item
the effect of applying the process denoted by a $t \in \CAPRAMProcTerm$ 
to a $\FlexVar^\RM_{\nm{deg}(t)}$-indexed data environment $\mu$ is 
$\eval{\rho}(t)$, where $\rho$ is a flexible variable valuation that 
represents $\mu$.
\end{itemize}

The \APRAMP\ model of computation described above is intended to be close 
to the asynchronous parallel RAM model of computation sketched 
in~\cite{CZ89a,KRS90a,Nis94a}.%
\footnote
{The \APRAMP\ model is considered less close to the asynchronous parallel
 RAM model sketched in~\cite{Gib89a} because the latter provides special
 instructions for synchronization. 
}
However, the time complexity measure for this model introduced in 
Section~\ref{sect-Measures} is quite different from the ones proposed 
in those papers.

The \APRAMP s can be looked upon as the processes that can be produced 
by a collection of BBRAMs with an extended instruction set when they 
execute their built-in program asynchronously in parallel.

The BBRAMs with the extended instruction set will be referred to as the 
SMBRAMs (Shared Memory Binary RAMs).
There is a strong resemblance between $\APRAMOp$ and the set $\APRAMInstr$
of instructions from which the built-in programs of the SMBRAMs can be 
constructed.
Because the concrete syntax of the instructions does not matter, 
$\APRAMInstr$ can be defined as follows:
\begin{ldispl}
\APRAMInstr = 
(\APRAMOp \Sdiff \RAMOpp) \Sunion
\Set{\jmpinstr{:}p{:}i \where p \in \RAMOpp \Land i \in \Natpos} \Sunion
\Set{\haltinstr}\;.
\end{ldispl}%
An \emph{SMBRAM program} is a non-empty sequence $C$ from $\APRAMProg^*$ 
in which instructions of the form $\jmpinstr{:}p{:}i$ with $i > \len(C)$ 
do not occur.
We write $\APRAMProg$ for the set of all SMBRAM programs.

For the SMBRAMs whose private memory has number $i$ ($i \in \Natpos$), 
the processes that are produced when they execute their built-in program 
are given by a function 
$\funct{\process_i}{\APRAMProg}{\APRAMProcTerm}$ that is defined up to
consistent renaming of variables as follows:
$\process_i(c_1\, \ldots\, c_n) = \rec{X_i}{E_i}$, 
where $E_i$ consists of the equation 
\begin{ldispl}
\begin{tabular}[t]{@{}l@{\,\,}l@{}}
$X_i = \True \gc \ass{\RM_i}{\iniop{:}\# i\,(\RM_i)} \seqc Y_1$
\end{tabular}
\end{ldispl}%
and, for each $j \in \Nat$ with $1 \leq j \leq n$, an equation
\begin{ldispl}
\begin{tabular}[t]{@{}l@{\,\,}l@{}}
$Y_j = \True \gc \ass{\RM_i}{c_j(\RM_i,\RM)} \seqc Y_{j+1}$ &
\hsp{9} if $c_j \in \nm{Load}$, 
\\
$Y_j = \True \gc \ass{\RM}{c_j(\RM_i,\RM)} \seqc Y_{j+1}$ &
\hsp{9} if $c_j \in \nm{Store}$, 
\\
$Y_j = \True \gc \ass{\RM_i}{c_j(\RM_i)} \seqc Y_{j+1}$ &
\hsp{9} if $c_j \in \RAMOpo$, 
\\
\multicolumn{2}{@{}l@{}}
{$Y_j = (p(\RM_i) = 1) \gc \ass{\RM_i}{\RM_i} \seqc Y_{j'}\altc 
        (p(\RM_i) = 0) \gc \ass{\RM_i}{\RM_i} \seqc Y_{j+1}$} \\ &
\hsp{9} if $c_j \equiv \jmpinstr{:}p{:}j'$, 
\\
$Y_j = \True \gc \ep$ &
\hsp{9} if $c_j \equiv \haltinstr$,
\end{tabular}
\end{ldispl}%
where 
$\nm{Load} =
 \Set{\loaop{:}@ i{:}\dst \where i \in \Nat \Land \dst \in \Dst}$,
$\nm{Store} =
 \Set{\stoop{:}\src{:}@ i \where \src \in Src \Land i \in \Nat}$, and
$Y_1,\ldots,Y_n$ are different variable from $\cX \Sdiff \Set{X_i}$.

The \APRAMP s are exactly the processes that can be produced by a 
collection of SMBRAMs when they execute their built-in program 
asynchronously in parallel.
\begin{theorem}
\label{theorem-APRAMP-SMRAM}
Let $n \in \Natpos$. 
For all constants 
$\rec{X_1}{E_1}, \ldots, \rec{X_n}{E_n} \in \ProcTermr$, 
$\rec{X_1}{E_1} \parc \ldots \parc \rec{X_n}{E_n} \in \APRAMProcTerm$ iff 
there exist $C_1, \ldots, C_n \in \APRAMProg$ such that 
$\rec{X_1}{E_1} \parc \ldots \parc \rec{X_n}{E_n}$ and 
$\process_1(C_1) \parc \ldots \parc \process_n(C_n)$ are identical up to 
consistent renaming of variables.
\end{theorem}
\begin{proof}
Let $i \in \Natpos$ be such that $i \leq n$.
It is easy to see that
(a)~for all $C \in \APRAMProg$, $\process_i(C) \in \APRAMProcTerm$ and 
(b)~$\process_i$~is an bijection up to consistent renaming of variables.
From this, it follows immediately that there exists a 
$C \in \APRAMProg$ such that $\rec{X_i}{E_i}$ and $\process_i(C)$ are 
identical up to consistent renaming of variables.
From this, the theorem follows immediately.
\qed
\end{proof}

\section{The \SPRAMP\ Model of Computation}
\label{sect-SPRAM-Model}

In the asynchronous parallel RAM model of computation presented in 
Section~\ref{sect-APRAM-Model}, the parallel processes that make up a 
computational process do not automatically synchronize after each 
computational step.
In this section, we describe a parallel RAM model of computation where
the parallel processes that make up a computational process 
automatically synchronize after each computational step.

For that purpose, a special instance of the synchronization merge 
operator of CSP~\cite{Hoa85} is defined in term of the operators of 
\deACPet+\RN.
It is assumed that $\sync,\synced \in \Act$ and $\commf$ is such that 
$\commf(\sync,\sync) = \synced$, 
$\commf(\sync,a) = \dead$ for all $a \in \Act \Sdiff \Set{\sync}$, and 
$\commf(\synced,a) = \dead$ for all $a \in \Act$.
The special instance of the synchronization merge operator, 
$\parc_\sync$, is defined as follows:
\begin{ldispl}
t \parc_\sync t' = 
\rnm{f}(\encap{\Set{\sync}}(\rnm{f}(t) \parc \rnm{f}(t')))\;,
\end{ldispl}%
where the renaming function $f$ is defined by $f(\synced) = \sync$ and 
$f(\alpha) = \alpha$ if $\alpha \in \AProcTerm \Sdiff \Set{\synced}$.

The parallel RAM model of computation described in this section is 
called the \SPRAMP\ (Synchronous Parallel Random Access Machine Process) 
model of computation. 

The operators that represent the RAM operations and the RAM properties 
that belong to $\gD$ in the case of the \SPRAMP\ model of computation
are the same as in the case of the \APRAMP\ model of computation.
The interpretation of those operators as a RAM operation or a RAM 
property is also the same as in the case of the \APRAMP\ model of 
computation.
Moreover, $\gD$ is fixed as in Section~\ref{sect-APRAM-Model}.

Below, the \SPRAMP\ model of computation is characterized.
However, first the notion of an $n$-\SPRAMP\ term is defined. 

Like in Section~\ref{sect-APRAM-Model}, it is assumed that 
$\RM \in \FlexVar$ and, for all $i \in \Natpos$, $\RM_i \in \FlexVar$.
Again, we write $\FlexVar^\RM_n$, where $n \in \Natpos$, for the set
$\Set{\RM} \Sunion \Set{\RM_i \where i \in \Natpos \Land i \leq n}$.

An \emph{$n$-SPRAM process term} ($n \in \Natpos$), called an 
\emph{$n$-\SPRAMP\ term} for short, is a term from $\ProcTermr$ that is 
of the form 
$\rec{X_1}{E_1} \parc_\sync \ldots \parc_\sync \rec{X_n}{E_n}$, where, 
for each $i \in \Natpos$ with $i \leq n$: 
\begin{itemize}
\item
for each $X \in \vars(E_i)$, 
the recursion equation for $X$ in $E_i$ has one of the following 
forms:
\begin{ldispl}
\begin{tabular}[t]{@{}l@{\;\;}l@{}}
(1) & $X = \True \gc \sync \seqc Y$, 
\\
(2) & $X = \True \gc \ass{\RM_i}{\iniop{:}\# i\,(\RM_i)} \seqc Y$, 
\\
(3) & $X = \True \gc \ass{\RM_i}{\loaop{:}@i{:}d\,(\RM_i,\RM)} \seqc Y$, 
\\
(4) & $X = \True \gc \ass{\RM}{\stoop{:}s{:}@i\,(\RM_i,\RM)} \seqc Y$, 
\\
(5) & $X = \True \gc \ass{\RM_i}{o(\RM_i)} \seqc Y$, 
\\
(6) & $X = (p(\RM_i) = 1) \gc \ass{\RM_i}{\RM_i} \seqc Y \altc 
       (p(\RM_i) = 0) \gc \ass{\RM_i}{\RM_i} \seqc Y'$, 
\\
(7) & $X = \True \gc \ep$,
\end{tabular}
\end{ldispl}%
where $o \in \RAMOpo$, $p \in \RAMOpp$, and 
$Y,Y' \in \vars(E_i)$;
\item
for each $X,Y \in \vars(E_i)$ with $Y$ occurring in the right-hand side 
of the recursion equation for $X$ in $E_i$,
the recursion equation for $X$ in $E_i$ is of the form~(1) iff
the recursion equation for $Y$ in $E_i$ is not of the form~(1);
\item
for each $X \in \vars(E_i)$, the recursion equation for $X$ in $E_i$ is 
of the form~(2) iff $X \equiv X_i$.
\end{itemize}
We write $\SPRAMProcTerm$ for the set of all terms $t \in \ProcTermr$
such that $t$ is an $n$-\SPRAMP\ terms for some $n \in \Natpos$, and
we write $\CSPRAMProcTerm$ for $\SPRAMProcTerm \Sinter \CProcTermr$.
Moreover, we write $\nm{deg}(t)$, where $t \in \SPRAMProcTerm$, for the 
unique $n \in \Natpos$ such that $t$ is an $n$-\SPRAMP\ term.

The terms from $\SPRAMProcTerm$ will be referred to as 
\emph{\SPRAMP\ terms}.

A process that can be denoted by an \SPRAMP\ term is called an
\emph{SPRAM process} or an \emph{\SPRAMP}\ for short.
So, an \SPRAMP\ is a synchronous parallel composition of processes that 
are definable by a guarded linear recursive specification over \deACPet\ 
of the kind described above.
Each of those parallel processes starts with an initialization step in 
which the number of its private memory is made available in the register 
with number $0$ from its private memory.

Notice that by Lemma~\ref{lemma-glr-abstr-free} and
Theorem~\ref{theorem-completeness-ACPet-1},
for all $t \in \SPRAMProcTerm$, there exists a guarded linear recursive 
specification $E$ and $X \in \vars(E)$ such that 
$t \rbbisim \rec{X}{E}$.

$\gD$ as fixed above and $\CSPRAMProcTerm$ induce the \SPRAMP\ model of 
computation: 
\begin{itemize}
\item
the set of possible computational processes is the set of all processes
that can be denoted by a term from $\CSPRAMProcTerm$;
\item
for each possible computational process $p$, the set of possible data 
environments is the set of all $\FlexVar^\RM_{\nm{deg}(t)}$-indexed data 
environments, where $t$ is a term from $\CSPRAMProcTerm$ denoting $p$; 
\item
the effect of applying the process denoted by a $t \in \CSPRAMProcTerm$ 
to a $\FlexVar^\RM_{\nm{deg}(t)}$-indexed data environment $\mu$ is 
$\eval{\rho}(t)$, where $\rho$ is a flexible variable valuation that 
represents $\mu$.
\end{itemize}

The \SPRAMP\ model of computation described above is intended to be 
close to the synchronous parallel RAM model of computation sketched 
\pagebreak[2]
in~\cite{SV84a}.%
\footnote
{The model sketched in~\cite{SV84a} is similar to, among others, the
 models sketched in~\cite{FW78a,Gol82a,KRS90a,TV94a,Mak97a}.
}
However, that model is a PRIORITY CRCW model whereas the \SPRAMP\ model
is essentially an ARBITRARY CRCW model (see~\cite{KR90a,Har94a}).
This means basically that, in the case that two or more parallel 
processes attempt to change the content of the same register at the same 
time, the process that succeeds in its attempt is chosen arbitrarily.
Moreover, in the model sketched in~\cite{SV84a}, the built-in programs 
of the RAMs that make up a PRAM must be the same whereas the parallel 
processes that make up an \SPRAMP\ may be different.

The \SPRAMP s can be looked upon as the processes that can be produced by 
a collection of SMBRAMs when they execute their built-in program 
synchronously in parallel.

For the SMBRAMs whose private memory has number $i$ ($i \in \Natpos$), 
the processes that are produced when they execute their built-in program 
are now given by a function 
$\funct{\process^\sync_i}{\APRAMProg}{\SPRAMProcTerm}$ that is defined 
up to consistent renaming of variables as follows:
$\process^\sync_i(c_1\, \ldots\, c_n) = \rec{X_i}{E_i}$, 
where $E_i$ consists of the equation 
\begin{ldispl}
\begin{tabular}[t]{@{}l@{\,\,}l@{}}
$X_i = \True \gc \ass{\RM_i}{\iniop{:}\# i\,(\RM_i)} \seqc Y_1$
\end{tabular}
\end{ldispl}%
and, for each $j \in \Nat$ with $1 \leq j \leq n$, an equation
\begin{ldispl}
\begin{tabular}[t]{@{}l@{\,\,}l@{}}
$Y_{2j-1} = \True \gc \sync \seqc Y_{2j}$
\\
$Y_{2j} = \True \gc \ass{\RM_i}{c_j(\RM_i,\RM)} \seqc Y_{2j+1}$ &
\hsp{8.6} if $c_j \in \nm{Load}$, 
\\
$Y_{2j} = \True \gc \ass{\RM}{c_j(\RM_i,\RM)} \seqc Y_{2j+1}$ &
\hsp{8.6} if $c_j \in \nm{Store}$, 
\\
$Y_{2j} = \True \gc \ass{\RM_i}{c_j(\RM_i)} \seqc Y_{2j+1}$ &
\hsp{8.6} if $c_j \in \RAMOpo$, 
\\
\multicolumn{2}{@{}l@{}}
{$Y_{2j} = (p(\RM_i) = 1) \gc \ass{\RM_i}{\RM_i} \seqc Y_{2j'-1}\altc 
           (p(\RM_i) = 0) \gc \ass{\RM_i}{\RM_i} \seqc Y_{2j+1}$} \\ &
\hsp{8.6} if $c_j \equiv \jmpinstr{:}p{:}j'$, 
\\
$Y_{2j} = \True \gc \ep$ &
\hsp{8.6} if $c_j \equiv \haltinstr$,
\end{tabular}
\end{ldispl}%
where 
$\nm{Load} =
 \Set{\loaop{:}@ i{:}\dst \where i \in \Nat \Land \dst \in \Dst}$,
$\nm{Store} =
 \Set{\stoop{:}\src{:}@ i \where \src \in Src \Land i \in \Nat}$, and
$Y_1,\ldots,Y_{2n}$ are different variable from $\cX \Sdiff \Set{X_i}$.

The \SPRAMP s are exactly the processes that can be produced by a 
collection of SMBRAMs when they execute their built-in program 
synchronously in parallel.
\begin{theorem}
\label{theorem-SPRAMP-SMRAM}
Let $n \in \Natpos$. 
For all constants 
$\rec{X_1}{E_1}, \ldots, \rec{X_n}{E_n} \in \ProcTermr$, 
$\rec{X_1}{E_1} \parc_\sync \ldots \parc_\sync \rec{X_n}{E_n} \in
 \SPRAMProcTerm$ iff 
there exist $C_1, \ldots, C_n \in \APRAMProg$ such that 
$\rec{X_1}{E_1} \parc_\sync \ldots \parc_\sync \rec{X_n}{E_n}$ and 
$\process^\sync_1(C_1) \parc_\sync \ldots \parc_\sync
 \process^\sync_n(C_n)$ 
are identical up to consistent renaming of variables.
\end{theorem}
\begin{proof}
Let $i \in \Natpos$ be such that $i \leq n$.
It is easy to see that
(a)~for all $C \in \APRAMProg$, 
$\process^\sync_i(C) \in \SPRAMProcTerm$ and 
(b)~$\process^\sync_i$~is an bijection up to consistent renaming of 
variables.
From this, it follows immediately that there exists a 
$C \in \APRAMProg$ such that $\rec{X_i}{E_i}$ and 
$\process^\sync_i(C)$ are identical up to consistent renaming of 
variables.
From this, the theorem follows immediately.
\qed
\end{proof}

The first synchronous parallel RAM models of computation, e.g.\ the 
models proposed in~\cite{FW78a,Gol82a,SV84a}, are older than the first 
asynchronous parallel RAM models of computation, e.g.\ the models 
proposed in~\cite{CZ89a,KRS90a,Nis94a}.
It appears that the synchronous parallel RAM models have been primarily 
devised to be used in the area of computational complexity and that the 
asynchronous parallel RAM models have been primarily devised because the 
synchronous models were considered of restricted value in the area of 
algorithm efficiency.

\section{Time and Work Complexity Measures}
\label{sect-Measures}

This section concerns complexity measures for the models of computation 
presented in Sections~\ref{sect-RAM-Model}--\ref{sect-SPRAM-Model}.
Before the complexity measures in question are introduced, it is made 
precise in the current setting what a complexity measure is and what the 
complexity of a computable function from ${(\BitStr)}^n$ to $\BitStr$ 
under a given complexity measure is. 

Let $\CProcTerm \subseteq \CProcTermr$.
Then a \emph{complexity measure} for $\CProcTerm$ is a partial function  
$\pfunct{M}{\CProcTerm \Sx \Union_{m \in \Nat} (\BitStr)^m}{\Nat}$ such 
that, for all $t \in \CProcTerm$ and 
$(w_1,\ldots,w_n) \in \Union_{m \in \Nat} (\BitStr)^m$, 
$M(t,(w_1,\ldots,w_n))$ is defined iff $\eval{\rho_{w_1,\ldots,w_n}}(t)$ 
is a terminating process term.%
\footnote
{This notion of a complexity measure bears little resemblance to Blum's
 notion of a complexity measure~\cite{Blu67a}, but it is in accordance
 with Blum's notion.}

Let $\CProcTerm \subseteq \CProcTermr$ and
let $M$ be a complexity measure for $\CProcTerm$.
Let $n \in \Nat$ and 
let $\pfunct{F}{{(\BitStr)}^n}{\BitStr}$ be a computable function. 
Let $\funct{V}{\Nat}{\Nat}$. 
Then \emph{$F$ is of complexity $V$ under the complexity measure $M$} if
there exists a $t \in \CProcTerm$ such that: 
\begin{itemize}
\item
$t$ computes $F$;
\item
for all $w_1,\ldots,w_n \in \BitStr$ such that $F(w_1,\ldots,w_n)$ 
is defined:
\begin{ldispl}
M(t,(w_1,\ldots,w_n)) \leq V(\len(w_1) + \ldots + \len(w_n))\;. 
\end{ldispl}%
\end{itemize}

\subsubsection*{The \RAMP\ Model of Computation} \mbox{}\\[1.5ex]
Below, a time complexity measure and a work complexity measure for the 
RAMP model of computation are introduced.

The \emph{sequential uniform time measure} is the complexity measure
$\SUTM$ for $\CRAMProcTerm$ defined by
\begin{ldispl}
\nm{\SUTM}(t,(w_1,\ldots,w_n)) = \depth(\eval{\rho_{w_1,\ldots,w_n}}(t)) 
\end{ldispl}%
for all $t \in \CRAMProcTerm$ and 
$(w_1,\ldots,w_n) \in \Union_{m \in \Nat} (\BitStr)^m$ 
such that $\eval{\rho_{w_1,\ldots,w_n}}(t)$ is a terminating process 
term. 

The sequential uniform time measure is essentially the same as the 
uniform time complexity measure for the standard RAM model of 
computation (see e.g~\cite{AHU74a}).
It is an idealized time measure: the simplifying assumption is made that
a RAMP performs one step per time unit.
That is, this measure actually yields, for a given RAMP and a given data 
environment, the maximum number of steps that can be performed by the 
given RAMP before eventually halting in the case where the initial data 
environment is the given data environment.
However, the maximum number of steps can also be looked upon as the 
maximum amount of work.
This makes the sequential uniform time measure a very plausible work 
measure as well.

The \emph{sequential work measure} is the complexity measure $\SWM$ for 
$\CRAMProcTerm$ defined by 
\begin{ldispl}
\nm{\SWM}(t,(w_1,\ldots,w_n)) = \nm{\SUTM}(t,(w_1,\ldots,w_n))
\end{ldispl}%
for all $t \in \CRAMProcTerm$ and 
$(w_1,\ldots,w_n) \in \Union_{m \in \Nat} (\BitStr)^m$ such that 
$\eval{\rho_{w_1,\ldots,w_n}}(t)$ is a terminating process term. 

In the sequential case, it is in accordance with our intuition that the 
uniform time complexity measure coincides with the work complexity 
measure.
In the parallel case, this is not in accordance with our intuition: it 
is to be expected that the introduction of parallelism results in a 
reduction of the amount of time needed but not in a reduction of the 
amount of work needed.

\subsubsection*{The \APRAMP\ Model of Computation} \mbox{}\\[1.5ex]
Below, a time complexity measure and a work complexity measure for the 
\APRAMP\ model of computation are introduced.

The \emph{asynchronous parallel uniform time measure} is the complexity 
measure $\APUTM$ for $\CAPRAMProcTerm$ defined by
\begin{ldispl}
\nm{\APUTM}(t,(w_1,\ldots,w_n)) = 
\max 
\Set{\depth(\abstr{\overline{\!H_i\!}}(\eval{\rho_{w_1,\ldots,w_n}}(t)))
     \where 1 \leq i \leq \nm{deg}(t)}\;,
\end{ldispl}%
where $\overline{\!H_i\!}$ is the set of all $\alpha \in \AProcTerm$ in 
which $\RM_i$ does not occur, 
for all $t \in \CAPRAMProcTerm$ and 
$(w_1,\ldots,w_n) \in \Union_{m \in \Nat} (\BitStr)^m$ such that 
$\eval{\rho_{w_1,\ldots,w_n}}(t)$ is a terminating process term. 

In the above definition, $\abstr{\overline{\!H_i\!}}$ turns steps of the
process denoted by $\eval{\rho_{w_1,\ldots,w_n}}(t)$ that are not 
performed by the parallel process whose private memory is referred to by 
$\RM_i$ into silent steps.
Because $\depth$ does not count silent steps, 
$\depth(\abstr{\overline{\!H_i\!}}(\eval{\rho_{w_1,\ldots,w_n}}(t)))$ is 
the maximum number of steps that the parallel process whose private 
memory is referred to by $\RM_i$ can perform.

Hence, the asynchronous parallel uniform time measure yields, for a 
given \APRAMP\ and a given data environment, the maximum over all 
parallel processes that make up the given \APRAMP\ of the maximum number 
of steps that can be performed before eventually halting in the case 
where the initial data environment is the given data environment.
Because it yields the maximum number of steps that can be performed by 
one of the parallel processes that make up the given \APRAMP, the
asynchronous parallel uniform time measure differs from the asynchronous 
parallel work measure.

The \emph{asynchronous parallel work measure} is the complexity measure
$\APWM$ for $\CAPRAMProcTerm$ defined by 
\begin{ldispl}
\nm{\APWM}(t,(w_1,\ldots,w_n)) = \depth(\eval{\rho_{w_1,\ldots,w_n}}(t))
\end{ldispl}%
for all $t \in \CAPRAMProcTerm$ and
$(w_1,\ldots,w_n) \in \Union_{m \in \Nat} (\BitStr)^m$ such that 
$\eval{\rho_{w_1,\ldots,w_n}}(t)$ is a terminating process term. 

The sequential work measure and the asynchronous parallel work measure 
are such that comparison of complexities under these measures have some 
meaning: both concern the maximum number of steps that can be performed 
by a computational process.

Like all complexity measures introduced in this section, the 
asynchronous parallel uniform time measure introduced above is a 
worst-case complexity measure.
It is quite different from the parallel time complexity measures that 
have been proposed for the asynchronous parallel RAM model of 
computation sketched in~\cite{CZ89a,KRS90a,Nis94a}.
The round complexity measure is proposed as parallel time complexity 
measure in~\cite{CZ89a,KRS90a} and an expected time complexity measure 
is proposed as parallel time complexity measure in~\cite{Nis94a}.
Neither of those measures is a worst-case complexity measure:
the round complexity measure removes certain cases from consideration
and the expected time complexity measure is an average-case complexity 
measure.

It appears that the round complexity measure and the expected time 
complexity measure are more important to analysis of the efficiency of 
parallel algorithms whereas the asynchronous parallel time complexity 
measure introduced above is more important to analysis of the complexity 
of computational problems that are amenable to solution by a parallel
algorithm.
After all, the area of computational complexity is mostly concerned with 
worst-case complexity.

In~\cite{Nis94a}, the asynchronous parallel uniform time measure 
introduced above is explicitly rejected.
Consider the case where there exists an interleaving of the parallel 
processes that make up an \APRAMP\ that is close to performing all steps 
of each of the processes uninterrupted by steps of the others.
Then the interleaving concerned is not ruled out by synchronization 
(through the shared memory) and may even be enforced by synchronization.
So it may be likely or unlikely to occur.
Seen in that light, it is surprising why it is stated in~\cite{Nis94a} 
that such an interleaving has ``very low probability, yielding a 
sequential measure''.

\subsubsection*{The \SPRAMP\ Model of Computation} \mbox{}\\[1.5ex]
Below, a time complexity measure and a work complexity measure for the 
\SPRAMP\ model of computation are introduced.

The time complexity measure introduced below is essentially the same as 
the uniform time complexity measure that goes with the synchronous 
parallel RAM model of computation sketched in~\cite{SV84a} and similar 
models.

The \emph{synchronous parallel uniform time measure} is the complexity 
measure $\SPUTM$ for $\CSPRAMProcTerm$ defined by
\begin{ldispl}
\SPUTM(t,(w_1,\ldots,w_n)) = 
\depth(\abstr{\overline{\sync}}(\eval{\rho_{w_1,\ldots,w_n}}(t)))\;,
\end{ldispl}%
where $\overline{\sync} = \AProcTerm \Sdiff \Set{\sync}$,
for all $t \in \CSPRAMProcTerm$ and
$(w_1,\ldots,w_n) \in \Union_{m \in \Nat} (\BitStr)^m$ such that 
$\eval{\rho_{w_1,\ldots,w_n}}(t)$ is a terminating process term. 

In the above definition, $\abstr{\overline{\sync}}$ turns all steps of 
the process denoted by $\eval{\rho_{w_1,\ldots,w_n}}(t)$ other than 
synchronization steps, i.e.\ all computational steps, into silent steps.
Because $\depth$ does not count silent steps, 
$\depth(\abstr{\overline{\sync}}(\eval{\rho_{w_1,\ldots,w_n}}(t)))$ is 
the maximum number of synchronization steps that can be performed by the 
process denoted by $\eval{\rho_{w_1,\ldots,w_n}}(t)$ before eventually 
halting.

Hence, the synchronous parallel uniform time measure yields, for a given 
\SPRAMP\ and a given data environment, the maximum number of 
synchronization steps that can be performed by the given \SPRAMP\ before
eventually halting in the case where the initial data environment is the 
given data environment.
Because the parallel processes that make up the given \SPRAMP\ synchronize
after each computational step, the time between two consecutive 
synchronization steps can be considered one time unit.
Therefore, this measure is a plausible time measure. 
Clearly, the maximum number of synchronization steps that can be 
performed by the given \SPRAMP\ and the maximum number of computational 
steps that can be performed by the given \SPRAMP\ are separate numbers.
So the synchronous parallel uniform time measure differs from the 
synchronous parallel work measure.

The \emph{synchronous parallel work measure} is the complexity measure
$\SPWM$ for $\CSPRAMProcTerm$ defined by 
\begin{ldispl}
\nm{\SPWM}(t,(w_1,\ldots,w_n)) =
 \depth(\abstr{\sync}(\eval{\rho_{w_1,\ldots,w_n}}(t))) 
\end{ldispl}%
for all $t \in \CSPRAMProcTerm$ and
$(w_1,\ldots,w_n) \in \Union_{m \in \Nat} (\BitStr)^m$ such that 
$\eval{\rho_{w_1,\ldots,w_n}}(t)$ is a terminating process term.

The sequential work measure and the synchronous parallel work measure 
are such that comparison of complexities under these measures have some 
meaning: both concern the maximum number of computational steps that can 
be performed by a computational process.

Take an \SPRAMP\ and the \APRAMP\ which is the \SPRAMP\ without the 
automatic synchronization after each computational step.
Assume that at any stage the next step to be taken by any of the 
parallel processes that make up the \APRAMP\ does not depend on the steps
that have been taken by the other parallel processes.
Then the synchronous parallel time measure $\SPUTM$ yields for the 
\SPRAMP\ the same result as the asynchronous parallel time measure 
$\APUTM$ yields for the \APRAMP.

\section{\SPRAMP s and the Parallel Computation Thesis}
\label{sect-SPRAM-Model-more}

The \SPRAMP\ model of computation is a simple model based on an 
idealization of existing shared memory parallel machines that abstracts 
from synchronization overhead.
The synchronous parallel uniform time measure introduced for this model
is a simple, hardware independent, and worst-case complexity measure.

The question is whether the \SPRAMP\ model of computation is a 
reasonable model of parallel computation.
A model of parallel computation is generally considered reasonable if 
the parallel computation thesis holds.
In the current setting, this thesis can be phrased as follows:
the \emph{parallel computation thesis} holds for a model of computation
if, for each computable partial function from ${(\BitStr)}^n$ to 
$\BitStr$ ($n \in \Nat$), its complexity under the time complexity 
measure for that model is polynomially related to its complexity under 
the space complexity measure for the multi-tape Turing machine model of 
computation.

Before we answer the question whether the \SPRAMP\ model of computation 
is a reasonable model of parallel computation, we go into a
classification of synchronous parallel RAMs.
This classification is used later on in answering the question. 
Below, synchronous parallel RAMs will be called PRAMs for short.

First of all, PRAMs can be classified as PRAMs whose constituent RAMs 
may execute different programs or PRAMs whose constituent RAMs must 
execute the same program.
The former PRAMs are called MIMD PRAMs and the latter PRAMs are called 
SIMD PRAMs.

In~\cite[Section 2.1]{KR90a}, PRAMs are classified according to their 
restrictions on shared memory access as 
EREW (Exclusive-Read Exclusive-Write), 
CREW (Concur\-rent-Read Exclusive-Write) or 
CRCW (Concurrent-Read Concurrent-Write).
CRCW PRAMs are further classified according to their way of resolving
write conflicts as  
COMMON, where all values attempted to be written concurrently into the 
same shared register must be identical,
ARBITRARY, where one of the values attempted to be written concurrently 
into the same shared register is chosen arbitrarily, or
PRIORITY, where the RAMs making up the PRAM are numbered and, from all 
values attempted to be written concurrently into the same shared 
register, the one attempted to be written by the RAM with the lowest 
number is chosen. 

Below, the next two lemmas about the above classifications of PRAMs 
will be used to show that the parallel computation thesis holds for the 
\SPRAMP\ model of computation.
\begin{lemma}
\label{lemma-arbitrary-priority}
Assuming a fixed instruction set:
\begin{enumerate}
\item
MIMD PRIORITY CRCW PRAMs can be simulated by MIMD ARBITRARY CRCW PRAMs 
with the same number of RAMs and with the parallel time increased by a 
factor of $O(\log(p))$, where $p$ is the number of RAMs;
\item 
MIMD ARBITRARY CRCW PRAMs can be simulated by MIMD PRIORITY CRCW PRAMs 
with the same number of RAMs and the same parallel time.
\end{enumerate}
\end{lemma}
\begin{proof}
Assume a fixed instruction set.

Part~1.\,
It is shown in~\cite[Section 3.1]{KR90a} that MIMD PRIORITY CRCW PRAMs 
can be simulated by MIMD EREW PRAMs with the same number of RAMs and 
with the parallel time increased by only a factor of $O(\log(p))$, where 
$p$ is the number of RAMs.
It follows directly from the definitions concerned that MIMD EREW PRAMs 
can be simulated by MIMD ARBITRARY CRCW PRAMs with the same number of 
RAMs and the same parallel time (the programs involved can be executed
directly).
Hence, MIMD PRIORITY CRCW PRAMs can be simulated by MIMD ARBITRARY CRCW 
PRAMs with the same number of RAMs and with the parallel time increased 
by a factor of $O(\log(p))$, where $p$ is the number of RAMs.

Part~2.\,
It follows directly from the definitions concerned that MIMD ARBITRARY 
CRCW PRAMs can be simulated by MIMD PRIORITY CRCW PRAMs with the same 
number of RAMs and the same parallel time (the programs involved can be 
executed directly).
\qed
\end{proof}

\begin{lemma}
\label{lemma-mimd-simd}
Assuming a fixed instruction set:
\begin{enumerate}
\item
SIMD PRIORITY CRCW PRAMs can be simulated by MIMD PRIORITY CRCW PRAMs
with the same number of RAMs and with the same parallel time;
\item
MIMD PRIORITY CRCW PRAMs can be simulated by SIMD PRIORITY CRCW PRAMs
with the same number of RAMs and with the parallel time increased by a 
constant factor.
\end{enumerate}
\end{lemma}
\begin{proof}
Assume a fixed instruction set.

Part~1.\,
This follows directly from the definitions concerned (the programs 
involved can be executed directly).

Part~2.\,
This is a special case of Theorem~3 from~\cite{Wlo91a}.
\qed
\end{proof}

The next theorem expresses that the parallel computation thesis holds 
for the \SPRAMP\ model of computation.
\begin{theorem}
\label{theorem-pra-comp-thesis}
Let $\pfunct{F}{{(\BitStr)}^m}{\BitStr}$ for some $m \in \Nat$ be a
computable function and let $\funct{T,S}{\Nat}{\Nat}$.
Then:
\begin{itemize}
\item
if $F$ is of complexity $T(n)$ under the synchronous parallel time 
complexity measure $\SPUTM$ for the \SPRAMP\ model of computation, then
there exists a $k \in \Nat$ such that $F$ is of complexity $O({T(n)}^k)$ 
under the space complexity measure for the multi-tape Turing machine 
model of computation;
\item
if $F$ is of complexity $S(n)$ under the space complexity measure for 
the multi-tape Turing machine model of computation, then
there exists a $k \in \Nat$ such that $F$ is of complexity $O({S(n)}^k)$ 
under the synchronous parallel time complexity measure $\SPUTM$ for the 
\SPRAMP\ model of computation provided that $S(n) \geq \log(n)$ for all 
$n \in \Nat$.
\end{itemize}
\end{theorem}
\begin{proof}
In~\cite{Gol82a}, SIMDAGs are introduced. 
SIMDAGs are SIMD PRIORITY CRCW PRAMs with a subset of the instruction 
set of SMBRAMs as instruction set.
Because 
$\mathrm{DSPACE}(S(n)) \subseteq \mathrm{NSPACE}(S(n)) \subseteq
 \mathrm{DSPACE}({S(n)}^2)$,
the variant of the current theorem for the SIMDAG model of computation 
follows immediately from Theorems~2.1 and~2.2 from~\cite{Gol82a} under 
a constructibility assumption for $S(n)$.
However, the proofs of those theorems go through with the instruction 
set of SMBRAMs because none of the SMBRAM instructions builds bit 
strings that are more than $O(T(n))$ bits long in $T(n)$ time. 
Moreover, if we take forking variants of SIMDAGs with the instruction 
set of SMBRAMs (resembling the P-RAMs from~\cite{FW78a}), the 
constructibility assumption for $S(n)$ is not needed.
This can be shown in the same way as in the proof of Lemma~1a 
from~\cite{FW78a}.

In the rest of this proof, we write E-SIMDAG for a SIMDAG with the
instruction set of SMBRAMs and forking E-SIMDAG for a forking variant of
an E-SIMDAG.

The variant of the current theorem for the forking E-SIMDAG model of 
computation follows directly from the above-mentioned facts. 

Now forking E-SIMDAGs can be simulated by E-SIMDAGs with $O(p)$ number 
of SMBRAMs and with the parallel time increased by a factor of 
$O(\log(p))$, where $p$ is the number of SMBRAMs used by the forking 
E-SIMDAG concerned.
This is proved as in the proof of Lemma~2.1 from~\cite{Goo89a}.
The other way round, E-SIMDAGs can be simulated by forking E-SIMDAGs 
with eventually the same number of SMBRAMs and with the parallel time 
increased by $O(\log(p))$, where $p$ is the number of SMBRAMs of the 
E-SIMDAG concerned.
This is easy to see: before the programs of the $p$ SMBRAMs involved can 
be executed directly, the $p$ SMBRAMs must be created by forking and 
this can be done in $O(\log(p))$ time.
It follows immediately from these simulation results that time 
complexities on forking E-SIMDAGs are polynomially related to time 
complexities on E-SIMDAGs.

The variant of the current theorem for the E-SIMDAG model of computation 
follows directly from the variant of the current theorem for the forking 
E-SIMDAG model of computation and the above-mentioned polynomial
relationship. 
From this, the fact that E-SIMDAGs are actually SIMD PRIORITY CRCW 
PRAMs that are composed of SMBRAMs, Lemma~\ref{lemma-mimd-simd}, 
Lemma~\ref{lemma-arbitrary-priority}, and 
Theorem~\ref{theorem-SPRAMP-SMRAM}, the current theorem now follows 
directly.
\qed
\end{proof}

\section{Probabilistic Computation}
\label{sect-PrRAM-Model}

In this section, it is first made precise in the setting introduced in 
Sections~\ref{sect-deACPet} and~\ref{sect-Computation} what it means 
that a given process probabilistically computes a given partial function 
from $(\BitStr)^n$ to $\BitStr$ ($n \in \Nat$).
Thereafter a probabilistic RAM model of computation and complexity 
measures for it are described in the setting introduced in 
Sections~\ref{sect-deACPet} and~\ref{sect-Computation}.

\subsection{Probabilistically Computing Partial Functions from 
\protect$(\BitStr)^n$ to $\BitStr$}
\label{subsect-prob-computation}

In this section, which is strongly based on~\cite{Gil77a}, it is made 
precise what it means that a given process probabilistically computes a 
given partial function from $(\BitStr)^n$ to $\BitStr$ ($n \in \Nat$).

Recall that $\gD$ is assumed to satisfy the RAM conditions given in
Section~\ref{subsect-RAM-Conds}.
Like in Section~\ref{subsect-computability}, it is assumed that 
$\RM \in \FlexVar$.
Moreover, it is assumed that $\toss \in \Act$ and $\commf$ is such that 
$\commf(\toss,a) = \dead$ for all $a \in \Act$.
We write $\overline{\toss} = \AProcTerm \Sdiff \Set{\toss}$.

The basic action $\toss$ is used to model probabilistic choices.
This is possible by the assumptions made about $\toss$: in the process 
denoted by a term of the form $\toss \seqc t \altc \toss \seqc t'$ the 
choice to behave as the process denoted by $t$ or the process denoted 
by $t'$ is made independent of anything, like with tossing a coin.
This allows for assuming that the probability that the first process is 
chosen and the probability that the second process is chosen are both 
$\tfrac{1}{2}$.

In order to model probabilistic choices in \deACPetrf, the use of the 
basic action $\toss$ has to be restricted to the modeling of 
probabilistic choices.
This restriction is covered by the subset $\ProcTermrt$ of $\ProcTermr$ 
defined as follows:
$\ProcTermrt$ is the set of all $t \in \ProcTermr$ for which there 
exists a guarded linear recursive specification $E$ and $X \in \vars(E)$ 
such that $\deACPetrf \Ent t = \rec{X}{E}$ and, 
for all $X' = t' \;\in\; E$ in which $\toss$ occurs, $t'$ is of the form
$\True \gc \toss \seqc Y' \altc \True \gc \toss \seqc Z'$, where 
$Y',Z' \in \vars(E)$.
By Lemma~\ref{lemma-glr-abstr-free}, $\ProcTermrt$ is well-defined.

In order to make precise what it means that a given process 
probabilistically computes a given partial function from $(\BitStr)^n$ 
to $\BitStr$, first three auxiliary notions are defined.

For all $t \in \BTcf$, the \emph{set of execution sequences of $t$}, 
written $\xseqs(t)$, is the subset of $\BTcf$ defined by simultaneous 
induction on the structure of $t$ as follows:
\begin{itemize}
\item
$\dead \in \xseqs(\dead)$;
\item
$\ep \in \xseqs(\ep)$;
\item 
if $\alpha \in \AProcTermt$ and $t' \in \xseqs(t)$, then 
$\alpha \seqc t' \in \xseqs(\alpha \seqc t)$;
\item
if $t \in \BTcf$ and $t'' \in \xseqs(t')$, then 
$t'' \in \xseqs(t \altc t')$;
\item
if $t \in \BTcf$ and $t'' \in \xseqs(t')$, then 
$t'' \in \xseqs(t' \altc t)$.
\end{itemize}
Let $t \in \ProcTermrt$ and $t' \in \BTcf$ be such that 
$\deACPetrf \Ent t = t'$.
Then we write $\xseqs(t)$ for $\xseqs(t')$.

Let $t \in \ProcTermrt$ and
$(w_1,\ldots,w_n) \in \Union_{m \in \Nat} (\BitStr)^m$ be such that 
$\eval{\rho_{w_1,\ldots,w_n}}(t)$ is a terminating process term, and 
let $w \in \BitStr$. 
Then 
the \emph{set of computation sequences of $(t,(w_1,\ldots,w_n))$ that 
yield $w$}, written $\cseqs(t,(w_1,\ldots,w_n),w)$, is the set of all 
$t' \in \xseqs(\eval{\rho_{w_1,\ldots,w_n}}(t))$ for which there exists 
a $\sigma \in \RMState$ with $\sigma(0) = w$ such that
\begin{ldispl}
\deACPetrf \Ent 
\eval{\rho_{w_1,\ldots,w_n}}(t') = 
\eval{\rho_{w_1,\ldots,w_n}}(t' \seqc (\RM = \sigma \gc \ep))\;.
\end{ldispl}%
We write $\cseqs_{\leq n}(t,(w_1,\ldots,w_n),w)$ for  
$\Set{t' \in \cseqs(t,(w_1,\ldots,w_n),w) \where
 \depth(\abstr{\toss}(t')) \leq n}$.

Let $t \in \ProcTermrt$,
$(w_1,\ldots,w_n) \in \Union_{m \in \Nat} (\BitStr)^m$, and 
$w \in \BitStr$ be such that $\eval{\rho_{w_1,\ldots,w_n}}(t)$ is a 
terminating process term and 
\mbox{$\cseqs(t,(w_1,\ldots,w_n),w) \neq \emptyset$}. 
Then the \emph{least number of steps in which $(t,(w_1,\ldots,w_n))$ 
yields $w$ with a probability greater than $\tfrac{1}{2}$}, written 
$\steps(t,(w_1,\ldots,w_n),w)$, is the least $n \in \Nat$ such that
\begin{ldispl}
\sum_{t' \in \cseqs_{\leq n}(t,(w_1,\ldots,w_n),w)}
  2^{- \depth(\abstr{\overline{\toss}}(t'))} > \tfrac{1}{2}\;.
\end{ldispl}%

Let $t \in \ProcTermrt$,
let $n \in \Nat$, let $\pfunct{F}{{(\BitStr)}^n}{\BitStr}$, and
let $\funct{W}{\Nat}{\Nat}$. \linebreak[2]
Then $t$ \emph{probabilistically computes $F$ in $W$ steps} if
\begin{itemize}
\item
for all $w_1,\ldots,w_n \in \BitStr$ such that $F(w_1,\ldots,w_n)$ 
is defined: 
\begin{ldispl}
\eval{\rho_{w_1,\ldots,w_n}}(t)\; 
\mathrm{is\, a\, terminating\, process\, term}\; \mathrm{and}\; \\
\cseqs(t,(w_1,\ldots,w_n),F(w_1,\ldots,w_n)) \neq \emptyset\;,
\seqnsep 
\steps(t,(w_1,\ldots,w_n),F(w_1,\ldots,w_n)) \leq 
W(\len(w_1) + \ldots + \len(w_n))\;; 
\end{ldispl}%
\item
for all $w_1,\ldots,w_n \in \BitStr$ such that $F(w_1,\ldots,w_n)$ is
undefined:
\begin{ldispl}
\eval{\rho_{w_1,\ldots,w_n}}(t)\; 
\mathrm{is\, not\, a\, terminating\, process\, term}\; \mathrm{or}\; \\
\cseqs(t,(w_1,\ldots,w_n),F(w_1,\ldots,w_n)) = \emptyset\;. 
\end{ldispl}%
\end{itemize}
We say that $t$ \emph{probabilistically computes $F$} if there exists a 
$\funct{W}{\Nat}{\Nat}$ such that $t$ probabilistically computes $F$ in 
$W$ steps,
we say that $F$ \emph{is a probabilistically computable function} if 
there exists a $t \in \ProcTerm$ such that $t$ computes $F$, and
we say that $t$ \emph{is a probabilistic computational process} if 
there exists a $\pfunct{F}{{(\BitStr)}^n}{\BitStr}$ such that $t$ 
probabilistically computes $F$.

We write $\PCProcTermr$ for  
$\Set{t \in \ProcTermrt \where
 t \mathrm{\;is\; a\; probabilistic\; computational\; process}}$.

Let $t \in \ProcTermrt$, $\pfunct{F}{{(\BitStr)}^n}{\BitStr}$ 
($n \in \Nat$), and $w_1,\ldots,w_n \in \BitStr$.
Suppose that $t$ probabilistically computes $F$ and $F(w_1,\ldots,w_n)$ 
is defined.
Then, there may be no computation sequences of $(t,(w_1,\ldots,w_n))$ or
computation sequences of $(t,(w_1,\ldots,w_n))$ that do not yield 
$F(w_1,\ldots,w_n)$.
The probability that this occurs is
\begin{ldispl}
1 -
\sum_{t' \in \cseqs(t,(w_1,\ldots,w_n),F(w_1,\ldots,w_n))}
  2^{- \depth(\abstr{\overline{\toss}}(t'))}\;.
\end{ldispl}%
If this \emph{error probability} is bounded below $\tfrac{1}{2}$ for all 
$w_1,\ldots,w_n \in \BitStr$ for which $F(w_1,\ldots,w_n)$ is defined,
it can be made arbitrary small merely by repeating the computation a 
bounded number of times and taking the majority result. 
This observation justifies the following definition.  

Let $t \in \ProcTermrt$ and $\pfunct{F}{{(\BitStr)}^n}{\BitStr}$ 
($n \in \Nat$) be such that $t$ probabilistically computes $F$.
Then $t$ probabilistically computes $F$ with 
\emph{bounded error probability} if there exists a $c < \tfrac{1}{2}$ 
such that 
\begin{ldispl}
1 -
 \sum_{t' \in \cseqs(t,(w_1,\ldots,w_n),F(w_1,\ldots,w_n))}
  2^{- \depth(\abstr{\overline{\toss}}(t'))} \leq c
\end{ldispl}%
for all $w_1,\ldots,w_n \in \BitStr$ for which $F(w_1,\ldots,w_n)$ 
is defined.

\subsection{The \PrRAMP\ Model of Computation}
\label{subsect-PrRAMP-model}

In this section, a probabilistic RAM model of computation is described 
in the setting introduced in Sections~\ref{sect-deACPet}, 
\ref{subsect-RAM-Conds}, and~\ref{subsect-prob-computation}.
Because it focuses on the processes that are produces by probabilistic 
RAMs when they execute their built-in programs, the probabilistic RAM 
model of computation described in this section is called the \PrRAMP\ 
(Probabilistic Random Access Machine Process) model of computation. 


In the case of the \PrRAMP\ model of computation, the set of operators 
from $\sign_\gD$ that are interpreted in $\gD$ as RAM operation or RAM 
property is the set $\PrRAMOp$ defined as follows:
\begin{ldispl}
\begin{aeqns}
\PrRAMOp & = & \RAMOp\;.
\end{aeqns}
\end{ldispl}%
For each $o \in \PrRAMOp$, the interpretation of $o$ in $\gD$, written
$\Int{o}$, is as defined in the case of the \RAMP\ model of computation 
in Section~\ref{subsect-Interpretation-Operators}.

In this section, as to be expected, $\gD$ is fixed as in 
Section~\ref{subsect-RAM-processes} for the \RAMP\ model of computation.
Moreover, like in Section~\ref{subsect-RAM-processes}, it is assumed 
that $\RM \in \FlexVar$.

Below, the notion of a \PrRAMP\ term is defined. 
This notion makes precise what the set of possible computational 
processes is in the case of the \PrRAMP\ model of computation.

A \emph{PrRAM process term}, called a \emph{\PrRAMP\ term} for short, 
is a term from $\ProcTermr$ that is of the form $\rec{X}{E}$, where, for 
each $Y \in \vars(E)$, the recursion equation for $Y$ in $E$ has one of 
the following forms:
\begin{ldispl}
\begin{tabular}[t]{@{}l@{}}
$Y = \True \gc \toss \seqc Z \altc \True \gc \toss \seqc Z'$,
\\
$Y = \True \gc \ass{\RM}{o(\RM)} \seqc Z$, 
\\
$Y = (p(\RM) = 1) \gc \ass{\RM}{\RM} \seqc Z \altc 
     (p(\RM) = 0) \gc \ass{\RM}{\RM} \seqc Z'$, 
\\
$Y = \True \gc \ep$,
\end{tabular}
\end{ldispl}%
where $o \in \RAMOpo$, $p \in \RAMOpp$, and 
$Z,Z' \in \vars(E)$.
We write $\PrRAMProcTerm$ for the set of all \PrRAMP\ terms, and
we write $\CPrRAMProcTerm$ for $\PrRAMProcTerm \Sinter \CProcTermr$.

A process that can be denoted by a \PrRAMP\ term is called a
\emph{PrRAM process} or a \emph{\PrRAMP}\ for short.
So, a \PrRAMP\ is a process that is definable by a guarded linear 
recursive specification over \deACPet\ of the kind described above.

$\gD$ as fixed above and $\CPrRAMProcTerm$ induce the \PrRAMP\ model of 
computation: 
\begin{itemize}
\item
the set of possible computational processes is the set of all processes
that can be denoted by a term from $\CPrRAMProcTerm$;
\item
for each possible computational process, the set of possible data 
environments is the set of all $\Set{\RM}$-indexed data environments; 
\item
the effect of applying the process denoted by a $t \in \CPrRAMProcTerm$ 
to a $\Set{\RM}$-indexed data environment $\mu$ is $\eval{\rho}(t)$, 
where $\rho$ is a flexible variable valuation that represents $\mu$.
\end{itemize}

To the best of my knowledge, only rough sketches of probabilistic RAMs 
are given in the computer science literature.
The \PrRAMP\ model of computation described above is in a way based on 
the probabilistic RAMs sketched in~\cite{Rei84a}.

In line with that paper, a variant of BBRAMs (defined below) that allows 
for probabilistic choices to be made are considered probabilistic RAMs.
They will be referred to as the PrBRAMs (Probabilistic Binary RAMs).

There is a strong resemblance between $\PrRAMOp$ and the set $\PrRAMInstr$
of instructions from which the built-in programs of the PrBRAMs can be 
constructed.
Because the concrete syntax of the instructions does not matter, 
$\PrRAMInstr$ can be defined as follows:
\begin{ldispl}
\PrRAMInstr = 
\RAMInstr \Sunion \Set{\prjmpinstr{:}i \where i \in \Natpos}\;.
\end{ldispl}%
A \emph{PrBRAM program} is a non-empty sequence $C$ from $\PrRAMInstr^*$ 
in which instructions of the form $\jmpinstr{:}p{:}i$ or the form
$\prjmpinstr{:}i$ with $i > \len(C)$ do not occur.
We write $\PrRAMProg$ for the set of all PrBRAM programs.

The execution of an instruction of the form $\prjmpinstr{:}i$ by a 
PrBRAM has no effect on the state of its memory.
After execution of an instruction of the form $\prjmpinstr{:}i$ by a 
PrBRAM, the execution proceeds with probability $\tfrac{1}{2}$ to the 
$i$th instruction of its built-in program and with probability 
$\tfrac{1}{2}$ to the next instruction from its built-in program.

The processes that are produced by the PrBRAMs when they execute their 
built-in program are given by a function 
$\funct{\process}{\PrRAMProg}{\PrRAMProcTerm}$ that is defined up to
consistent renaming of variables as follows:
$\process(c_1\, \ldots\, c_n) = \rec{X_1}{E}$, 
where \linebreak[2] $E$ consists of, 
for each $i \in \Nat$ with $1 \leq i \leq n$, an equation
\begin{ldispl}
\begin{tabular}[t]{@{}l@{\,\,}l@{}}
$X_i = \True \gc \toss \seqc X_j\altc 
        \True \gc \toss \seqc X_{i+1}$ &
\hsp{9} if $c_i \equiv \prjmpinstr{:}j$, 
\\
$X_i = \True \gc \ass{\RM}{c_i(\RM)} \seqc X_{i+1}$ &
\hsp{9} if $c_i \in \RAMOpo$, 
\\
\multicolumn{2}{@{}l@{}}
{$X_i = (p(\RM) = 1) \gc \ass{\RM}{\RM} \seqc X_j\altc 
       (p(\RM) = 0) \gc \ass{\RM}{\RM} \seqc X_{i+1}$} \\ &
\hsp{9} if $c_i \equiv \jmpinstr{:}p{:}j$, 
\\
$X_i = \True \gc \ep$ &
\hsp{9} if $c_i \equiv \haltinstr$,
\end{tabular}
\end{ldispl}%
where $X_1,\ldots,X_n$ are different variable from $\cX$.

Let $C \in \PrRAMProg$.
Then $\process(C)$ denotes the process that is produced by the PrBRAM 
whose built-in program is $C$ when it executes its built-in program.

The definition of $\process$ is in accordance with the sketch of the 
version of the probabilistic RAM model of computation in~\cite{Rei84a}.
However, that sketch is not precise and complete enough to allow of a 
proof of this.

The PrRAMPs are exactly the processes that can be produced by the 
PrBRAMs when they execute their built-in program.
\begin{theorem}
\label{theorem-PrRAMP-PrRAM}
For each constant $\rec{X}{E} \in \ProcTermr$, 
$\rec{X}{E} \in \PrRAMProcTerm$ iff 
there exists a $C \in \PrRAMProg$ such that $\rec{X}{E}$ and 
$\process(C)$ are identical up to consistent renaming of variables.
\end{theorem}
\begin{proof}
It is easy to see that
(a)~for all $C \in \PrRAMProg$, $\process(C) \in \PrRAMProcTerm$ and 
(b)~$\process$~is an bijection up to consistent renaming of variables.
From this, the theorem follows immediately.
\qed
\end{proof}

The following theorem is a result concerning the computational power of
PrRAMPs.
\begin{theorem}
\label{theorem-Turing-computable-PrRAMP}
For each $\pfunct{F}{{(\BitStr)}^n}{\BitStr}$, there exists a 
$t \in \PrRAMProcTerm$ such that $t$ probabilistically computes $F$ iff 
$F$ is Turing-computable.
\end{theorem}
\begin{proof}
By Theorem~\ref{theorem-Turing-computable-RAMP}, the proof goes like the 
proof of Proposition~2.3 in~\cite{Gil77a}.
\qed
\end{proof}

In~\cite{Gil77a}, the complexity classes $\mathbf{PP}$ and 
$\mathbf{BPP}$ are defined in terms of probabilistic Turing machines.
Theorem~\ref{theorem-PP-and-BPP} shows that those complexity classes can
also be defined in terms of PrRAMPs.
\begin{theorem}
\label{theorem-PP-and-BPP}
For each $\funct{F}{\BitStr}{\Bit}$: \vspace*{-1ex} 
\begin{itemize}
\item
there exist a $t \in \PrRAMProcTerm$ and a $W \in \POLY$ such that  
$t$ probabilistically computes $F$ in $W$ steps iff $F \in \mathbf{PP}$;
\item
there exist a $t \in \PrRAMProcTerm$ and a $W \in \POLY$ such that 
$t$ probabilistically computes $F$ in $W$ steps with bounded error 
probability iff $F \in \mathbf{BPP}$.
\end{itemize}
\end{theorem}
\begin{proof}
By Theorem~\ref{theorem-PrRAMP-PrRAM} and the definitions of 
$\mathbf{PP}$ and $\mathbf{BPP}$ referred to above, it is sufficient to 
prove that time complexity on PrBRAMs under the uniform time measure, 
i.e.\ the number of steps, and time complexity on multi-tape 
probabilistic Turing machines are polynomially related.
The proof goes like the proof of Theorem~2 in~\cite{CR73a} because the
following holds for each instruction from $\PrRAMInstr$:
(a)~the execution of the instruction affects at most one register and
(b)~the execution of the instruction increases the maximum of the number 
of bits in the registers at most by one.
\qed
\end{proof}

At the end of Section~\ref{subsect-prob-computation}, the error 
probability of probabilistic computational processes was shortly 
discussed.
The complexity class $\mathbf{BPP}$ is interesting because, for each 
$\funct{F}{\BitStr}{\Bit}$, the reduction of the error probability by 
repeating the computation is exponential in the number of repetitions 
if $F \in \mathbf{BPP}$ (see e.g.~\cite{AB09a}).
This means that an arbitrary reduction of the error probability is 
possible by repeating the computation a bounded number of times if 
$F \in \mathbf{BPP}$.
Because of that, $\mathbf{BPP}$ is nearly as feasible a complexity class 
as $\mathbf{P}$.
An arbitrary reduction is not possible by repeating the computation a 
bounded number of times if $F \in \mathbf{PP} \Sdiff \mathbf{BPP}$.

In the same vein as the probabilistic variant of the RAMP model of 
computation has been described above, probabilistic variants of the 
APRAMP model and the SPRAMP model can be described.

\subsection{Time and Work Complexity Measures}
\label{subsect-measures-PrRAMP-model}

Below, a probabilistic time complexity measure for the \PrRAMP\ model of 
computation is introduced.
In preparation, it is first made precise what a probabilistic complexity 
measure is and what the complexity of a probabilistically computable 
function from ${(\BitStr)}^n$ to $\BitStr$ under a given probabilistic 
complexity measure is. 
The notion of a complexity measure have to be adapted to the 
probabilistic case because the probabilistic computation of a function 
may not always yield the correct result.

Let $\PCProcTerm \subseteq \PCProcTermr$.
Then a \emph{probabilistic complexity measure} for $\PCProcTerm$ is a 
partial function 
$\pfunct{M}
  {\PCProcTerm \Sx \Union_{m \in \Nat} (\BitStr)^m \Sx \BitStr}{\Nat}$ 
such that, for all $t \in \PCProcTerm$, 
$(w_1,\ldots,w_n) \in \Union_{m \in \Nat} (\BitStr)^m$, and 
$w \in \BitStr$, $M(t,(w_1,\ldots,w_n),w)$ is defined iff 
$\eval{\rho_{w_1,\ldots,w_n}}(t)$ is a terminating process 
term and $\cseqs(t,(w_1,\ldots,w_n),w) \neq \emptyset$.

Let $\PCProcTerm \subseteq \PCProcTermr$ and
let $M$ be a probabilistic complexity measure for $\PCProcTerm$.
Let $n \in \Nat$ and 
let $\pfunct{F}{{(\BitStr)}^n}{\BitStr}$ be a probabilistically 
computable function. 
Let $\funct{V}{\Nat}{\Nat}$. 
Then \emph{$F$ is of complexity $V$ under the probabilistic complexity 
measure $M$} if there exists a $t \in \PCProcTerm$ such that: 
\begin{itemize}
\item
$t$ probabilistically computes $F$;
\item
for all $w_1,\ldots,w_n \in \BitStr$ such that $F(w_1,\ldots,w_n)$ 
is defined:
\begin{ldispl}
M(t,(w_1,\ldots,w_n),F(w_1,\ldots,w_n)) \leq 
V(\len(w_1) + \ldots + \len(w_n))\;. 
\end{ldispl}%
\end{itemize}

The \emph{probabilistic sequential uniform time measure} is the 
probabilistic complexity measure $\PrSUTM$ for $\CPrRAMProcTerm$ defined 
by
\begin{ldispl}
\nm{\PrSUTM}(t,(w_1,\ldots,w_n),w) = \steps(t,(w_1,\ldots,w_n),w) 
\end{ldispl}%
for all $t \in \CPrRAMProcTerm$, 
$(w_1,\ldots,w_n) \in \Union_{m \in \Nat} (\BitStr)^m$, and 
$w \in \BitStr$ such that $\eval{\rho_{w_1,\ldots,w_n}}(t)$ is a 
terminating process term and 
$\cseqs(t,(w_1,\ldots,w_n),w) \neq \emptyset$. 

Like for the RAMP model of computation, the probabilistic sequential 
uniform time measure a very plausible work measure as well.

The \emph{probabilistic sequential work measure} is the probabilistic 
complexity measure $\PrSUTM$ for $\CPrRAMProcTerm$ defined by
\begin{ldispl}
\nm{\PrSWM}(t,(w_1,\ldots,w_n),w) = \nm{\PrSUTM}(t,(w_1,\ldots,w_n),w) 
\end{ldispl}%
for all $t \in \CPrRAMProcTerm$, 
$(w_1,\ldots,w_n) \in \Union_{m \in \Nat} (\BitStr)^m$, and 
$w \in \BitStr$ such that $\eval{\rho_{w_1,\ldots,w_n}}(t)$ is a 
terminating process term and 
$\cseqs(t,(w_1,\ldots,w_n),w) \neq \emptyset$. 

For the probabilistic variants of the APRAMP model and the SPRAMP model, 
the definitions of the suitable variants of the complexity measures that 
have been introduced for the APRAMP model and the SPRAMP model are 
somewhat more involved.

\section{Concluding Remarks}
\label{sect-conclusions}

In this paper, it has been studied whether the imperative process algebra 
\deACPet\ \linebreak[2] can play a role in the field of models of 
computation. 
Models of computation corresponding to models based on sequential random 
access machines, asynchronous parallel random access machines, 
synchronous parallel random access machines, a probabilistic variant of 
sequential random access machines, and complexity measures for those 
models could simply and directly be described in a mathematically 
precise way in the setting of \deACPet.
A probabilistic variant of the model based on sequential random access 
machines and complexity measures for it could also be described, but in 
a somewhat less simple and direct way. \linebreak[2]
Central in the models described are the computational processes 
considered instead of the abstract machines that produce those processes
when they execute their built-in program.

The work presented in this paper pertains to formalizing models of
computation.
Little work has been done in this area.
Three notable exceptions are~\cite{Nor11a,XZU13a,AR15a}.
Those papers are concerned with formalization in a theorem prover (HOL4, 
Isabelle/HOL, Matita) and focusses on some version of the Turing machine 
model of computation.
This makes it impracticable to compare the work presented in those 
papers with the work presented in this paper.

Whereas it is usual in versions of the RAM model of computation that bit 
strings are represented by natural numbers, here natural numbers are 
represented by bit strings. 
Moreover, the choice has been made to represent the natural number $0$ 
by the bit string $0$ and to adopt the empty bit string as the register 
content that indicates that a register is (as yet) unused.


\bibliographystyle{splncs03}
\bibliography{PA}

\end{document}